\documentclass[pdflatex,12pt]{article}
\usepackage{url}
\usepackage{jheppub} 
\usepackage{fancyhdr}
\usepackage{xcolor}
\usepackage{graphicx}
\usepackage{rotating}
\usepackage{comment}
\usepackage{braket}
\usepackage{placeins}
\definecolor{darkgreen}{rgb}{0,0.5,0}
\definecolor{darkblue}{rgb}{0,0,0.6}
\definecolor{darkred}{rgb}{0.6,0,0.0}
\newcommand{\p}{\partial}
\newcommand{\mn}{{\mu\nu}}
\newcommand{\ra}{\rightarrow}
\newcommand{\ali}[1]{\begin{align} #1 \end{align}}

\newcommand{\ph}[1]{\phantom{#1}}

\newcommand{\U}{\mathsf{U}}
\newcommand{\SL}{\mathsf{SL}}
\newcommand{\R}{\mathbb{R}}
\newcommand{\pd}{\,\partial}
\newcommand{\br}[1]{\left[#1\right]}
\newcommand{\pa}[1]{\left(#1\right)}

\numberwithin{equation}{section}
\allowdisplaybreaks  

\title{Electro-Magnetic Energy Extraction from Rotating Black Holes in AdS}

\author[a,b]{Nele Callebaut,}
\author[c,d]{Maria J. Rodriguez\,}
\author[e]{and Herman Verlinde\,}

\affiliation[a]{Department of Mathematics and 
	Haifa Research Center for Theoretical Physics and Astrophysics, 
	University of Haifa, Haifa 31905, Israel} 
\affiliation[b]{Department of Physics, Technion,
	Haifa 32000, Israel}
\affiliation[c]{ Department of Physics, Utah State University, 4415 Old Main Hill Road, UT 84322, USA}
\affiliation[d]{  Instituto de F\' isica Te\' orica UAM/CSIC, Universidad Aut\'  onoma de Madrid, Cantoblanco, 28049 Madrid, Spain} 
\affiliation[e]{Joseph Henry Laboratories, Princeton University, Princeton, NJ 08544, USA} 

\emailAdd{nele\_calleba@technion.ac.il} 
\emailAdd{majo.rodriguez.b@gmail.com} 
\emailAdd{verlinde@princeton.edu}

\abstract{
	Force-Free Electrodynamics for black holes in Anti de Sitter is considered. We present new, energy extracting solutions of Force-Free Electrodynamics in Anti de Sitter - Near Horizon Extremal Kerr and Super-Entropic Near Horizon Extremal Kerr geometries. The relevant equations of motion are derived from an action for force-free plasma surrounding spinning black holes with generic asymptotics. We consider the energy flux of electrodynamic fields in rotating frames to argue that the correct measure for energy extraction is the energy flux measured by a rotating observer in the near horizon region. We illustrate this procedure by application to near horizon solutions in Kerr, AdS-Kerr and BTZ. 
}

\begin{document} 
	\maketitle

	
\section{Introduction}

It is widely believed that astrophysical black holes are typically surrounded by magnetospheres composed of an electromagnetic
plasma that is force-free to a good approximation. 
In \cite{Blandford:1977ds}, Blandford and Znajek used a model of force-free electrodynamics (FFE) for the plasma surrounding a spinning black hole to suggest a mechanism for energy extraction from the black hole. They argued that rotational energy escapes via plasma currents 
along magnetic field lines that thread the horizon. The  Blandford Znajek (BZ) process is considered to provide the basic picture 
behind astrophysical observations of jets, e.g.~in active galactic nuclei and quasars. 
For more recent discussions on the consistency  of the BZ mechanism, see \cite{Grignani:2018ntq,Grignani:2019dqc,Armas:2020mio}. 
	
	The FFE solution discussed in the original paper  \cite{Blandford:1977ds} is an approximate analytical solution for a slowly rotating Kerr black hole. Despite a lot of work on numerical as well as analytical solutions (\cite{1973MichelCurtis,1976Blandford,Menon:2005mg,Tanabe:2008wm,Brennan:2013kea,Gralla:2015vta}, see also \cite{Gralla:2014yja} and references therein), it remains a challenge to find a physically realistic, exact and energy extracting solution of the FFE equations in a Kerr black hole background. 
	
	One successful strategy for generating families of solutions has been to exploit the enhanced amount of symmetry in the near-horizon region of an extremal, i.e.~maximally spinning,  Kerr black hole \cite{Lupsasca:2014pfa,Lupsasca:2014hua,Zhang:2014pla}. In this paper we will continue this effort by proposing a previously overlooked symmetric ansatz for the electromagnetic field strength of the plasma. We discuss in detail the corresponding energy outflux. By making use of a toy model of a rotating electromagnetic configuration, we argue the correct measure for energy extraction is the outflux measured by a \emph{rotating} observer in the near-horizon geometry, corresponding to an observer at rest in Kerr. 
	The new symmetric ansatz has the main advantage (over previous energy-extracting, symmetric FFE solutions \cite{Zhang:2014pla}) that it gives rise to an energy outflux at the boundary of the near-horizon region, as measured by a Kerr observer, that is finite.     
	
	The FFE solutions that can be produced with the new ansatz in the Near Horizon Extremal Kerr geometry -- known as NHEK -- do not obey the assumed physical boundary conditions presented in this paper. However, when we generalize the background to the Near Horizon Extremal \emph{AdS}-Kerr geometry -- introduced later as AdS-NHEK or SE-NHEK for super-entropic Near Horizon Extremal AdS-Kerr -- we do find such a solution. For the background in question the topology is in fact that of a cylinder and this metric therefore provides a close resemblance to our toy model of a rotating conducting cylinder.    
	
	The generalization to AdS-Kerr was motivated by the discovery in \cite{Jacobson:2017xam} of an exotic application of the BZ process that does not involve the presence of plasma, realized in a BTZ background.    
	While not immediately relevant astrophysically, the study of FFE in AdS backgrounds may prove useful for obtaining a better understanding of remaining conceptual questions regarding the BZ process, such as where and how the negative energy originates. Indeed we hope the presented toy model and FFE solutions in near-horizon AdS-Kerr may contribute to this goal in the future.  \\
	
	The toy model and new symmetric log ansatz in NHEK were first presented in the unpublished \cite{ThesisLibby}. In this paper we additionally apply these ideas in the context of AdS-Kerr black holes.    \\

	The paper is organized as follows. We begin in section \ref{sectionaction} with a review where we explain how a 2-dimensional (2D) action naturally describes 
	toroidal force-free plasma systems surrounding a spinning black hole. At this stage we already assume a system that is toroidally invariant and therefore effectively 2-dimensional. 
	We present in section \ref{secNearHorizons} all the near-horizon geometries of spinning black holes that we will be using in the paper. Next, in section \ref{sectionFFEsols}, we apply the technique of imposing the extra scaling symmetry of the near-horizon regions on the field strength of the plasma. This gives rise to the scaling ansatz in \ref{sectionscaling} and the new log ansatz in \ref{sectionlogansatz}. These ansatze fix the analytic behavior of the field strength as a function of the radius $r$ of the background.  
	The scaling ansatz is known in the literature for NHEK. We employ the same scaling ansatz in AdS-NHEK and SE-NHEK to find a new scaling solution of FFE. The log ansatz, to our knowledge, is new in all three backgrounds. 
	
	Making use of the ansatze, the problem reduces to solving an ODE for a function $\Phi(\theta)$ describing the dependence of the field strength on the polar angle. This is the part that will be treated numerically. The numerical problem is to solve the EOM summarized in section \ref{sectionEOM} with physical boundary conditions provided in section \ref{subsec:bc}. 

	Before presenting the numerical solutions, in section \ref{sectionenergy} we consider the energy flux of electrodynamic fields in rotating frames, making use of a `toy model', to explain how the fluxes at the boundary of the near-horizon throats connect with non-rotating observers outside the throat. We argue it is the latter observer's flux that determines whether a near-horizon solution is energy extracting. The main results of this section are equation \eqref{toymodel49} for the energy flux in the toy model, equation \eqref{ErKerr} for the Kerr energy flux and \eqref{ErKerrAdS} for the AdS-Kerr energy flux. 
	The general relation \eqref{fluxsign} highlights the equivalence with the toy model. 
These formulas are applied to our energy-extracting semi-analytical solutions in section \ref{sectionwithfigs}.  
The energy extraction is also discussed in section \ref{subsec:BTZ} for the full BTZ solution of \cite{Jacobson:2017xam}. This provides a check on the proposed procedure for obtaining the flux outside the near-horizon throat.  
	
Finally we present a scaling solution in NHEK, a scaling solution in AdS-NHEK and a log solution in SE-NHEK, with their corresponding Poynting fluxes, in Figures \ref{figscalingNHEK}-\ref{figSENHEKlogII}.   Final remarks are discussed in section \ref{sectiondiscussion}. Some mathematical details of our analysis have been relegated to Appendices \ref{appKerr}-\ref{appCurvedEM}. \\

We adopt the metric signature $(-,+,+,+)$, the units $c=G=1$, and orientation of $dt \wedge d\phi \wedge dr \wedge d\theta$ for defining the Hodge dual $\star$.


\section{Action for force-free plasma in black holes} 
\label{sectionaction} 

In this section we review  the set-up of the problem, hereby setting our notation. In particular, we will make use of an available action principle for the description of a force-free plasma in black holes. This has several advantages in general, such as allowing for the easy identification of conserved quantities through Noether's theorem.  

As first studied 
by Blandford and Znajek \cite{Blandford:1977ds}, the equations for the force-free plasma around rotating black holes (described by a fixed background metric $g_{\mu\nu}$) consists of Maxwell's equations 
\begin{eqnarray}
\nabla_\mu F^{\nu\mu} &=& J^\nu\,,   \label{eom} \\
\nabla_{[\mu} F_{\sigma\nu]} &=& 0\,,  \label{bianchi}
\end{eqnarray} 
and the force-free electromagnetic (FFE)  condition
\begin{equation}\label{eq:ffe}
F_{\mu\nu} J^\nu = 0\,,
\end{equation}
with source current $J_\nu$ and field strength $F_{\mu\nu}$. Solutions to these equations involve finding a field strength $F_{\mu\nu}$ that generates a current $J^{\mu}$ through (\ref{eom}), and satisfies the FFE constraint (\ref{eq:ffe}). The Bianchi identity \eqref{bianchi} follows straightforwardly by writing the field strength as $F_{\mu\nu} = \partial_\mu A_\nu - \partial_\nu A_\mu$ in terms of the gauge field $A_\mu$.

The criterion for the FFE condition (\ref{eq:ffe}) to hold is that in local inertial frames the matter source  contribution $T^{matter}_{\mu\nu}$ becomes negligible.
The stress tensor can therefore be defined as
\begin{equation}\label{FFEapprox}
T_{\mu\nu}=T^{EM}_{\mu\nu}+T^{matter}_{\mu\nu} \approx  T_{\mu\nu}^{EM}\,,
\end{equation}
where $T_{\mu\nu}^{EM}$ is the contribution from the electric and magnetic field.\footnote{The correspondence with the force-free condition $(\ref{eq:ffe})$ follows directly from stress-energy conservation $\nabla_\mu  T^{\mu\nu}=0$, combined with $\nabla^\nu T^{EM}_{\mu\nu}=-F_{\mu\nu}J^{\nu}$ from Maxwell's equations. } In this FFE regime, the Einstein-Maxwell action for the interaction between the charged matter and the gauge field $A_{\mu}$ must be nearly independent of $g_{\mu\nu}$, hence
\begin{equation}\label{eq:action}
\mathcal{S}[g,A]\rightarrow \mathcal{S}[A] =\int \sqrt{-g} \left( -\frac{1}{4} F_{\mu\nu}F^{\mu\nu}+ A_{\mu} J^{\mu}\right) dt \,  d\phi   \, dr  \, d\theta\,.
\end{equation} 
Maxwell's inhomogeneous equations $(\ref{eom})$ follow as the Euler-Lagrange equations of this action.

In general, the above system of equations \eqref{eom}-\eqref{eq:ffe} 
is highly nonlinear and can only be solved numerically. However, in toroidal spacetimes the symmetries can be exploited to simplify the analysis and obtain semi-analytical solutions  \cite{Lupsasca:2014pfa,Lupsasca:2014hua,Zhang:2014pla}. Here, by toroidal spacetimes we mean curved spacetimes that can be described by a coordinate system $(t,\phi,r,\theta)$ in which the metric is independent of time $t$ and angle $\phi$. 
The coordinates $(t,\phi)$ are then referred to as {\it toroidal} coordinates $x^{\alpha,\beta}$, and $(r,\theta)$ as {\it poloidal} coordinates $x^{a,b}$. The line-element of such a spacetime with a block-diagonal metric ($g_{a\alpha} = 0$), independent of time $t$ and angle $\phi$ ($\p_t g_\mn =  \p_\phi g_\mn$=0) takes the form 
\begin{equation}\label{metric0}
ds^2\equiv (g_T)_{\alpha\beta}\,dx^{\alpha}dx^{\beta}+(g_P)_{ab}\,dx^{a}dx^{b}\, , 
\end{equation} 
with toroidal metric $g_T$ and poloidal metric $g_P$. 

Typically the dynamics around a black hole will capture the black hole's symmetries. It is natural then to impose the same toroidal symmetry of the background metric on the solutions for the field strength. As we now argue, this will allow to reduce the 4D problem to a well-defined action in the two poloidal coordinates. 

We assume a stationary and axisymmetric solution, characterized by a field strength with non-zero components 
\ali{
	F_{\mu a} (x^a) \neq 0\,. 
	} 
The analysis of the conditions on $F_{\mu a}$ imposed by the FFE equations is most clear in form notation, with the field strength defined as a two-form $F = \frac{1}{2} F_\mn dx^\mu \wedge dx^\nu$. See e.g.~\cite{Jacobson:2017xam} for a quick  review of differential forms in the context of force-free electrodynamics. In this notation, the equations \eqref{eom}-\eqref{bianchi} 
are written respectively as 
\begin{eqnarray}
 d \star F &=& \star J\,\label{Maxwell1}\\
 d F &=& 0\, 
\end{eqnarray}
for the 2-form field strength $F = dA$ in terms of the 1-form gauge field $A$, with $\star$ the Hodge dual and $\wedge$ the wedge product.
Introducing the notation $A_T=A_\alpha dx^\alpha$ for the toroidal contribution and $A_P=A_a dx^a$ for the poloidal contribution to the gauge field, we can write
\ali{
	A = A_\mu dx^\mu= A_T + A_P  
}
and 
\ali{
	F =dA_t \wedge dt + dA_\phi \wedge d\phi  + dA_P \,\,.  \label{F}
}
Similarly, the 1-form current $J$ is 
\ali{
	J = J_P + J_T \,\, . 
}
In the expression for the field strength, $d$ is the differential operator in 4D, with $dA = \p_\mu A \, dx^\mu$. Imposing toroidal invariance comes down to reinterpreting the  differential operator $d$ as a 2D differential operator in the poloidal metric $g_P$.  
That is, $dA = \p_a A \, dx^a$. The toroidal invariance is therefore implicitly present in the expressions through the interpretation of $d$.   

The toroidal component of the FFE equation \eqref{eq:ffe} can be written in form notation as 
\ali{
	dA_T \wedge *J_P = 0 \,\,.  \label{FFEform}
}
Here, we use $*$ to refer to the star operator in the poloidal metric,  to be distinguished from the star operator $\star$ in the full 4D spacetime (as used in Maxwell's equation). Its action is defined as $* J_P = J_a *dx^a = J_a \epsilon^a_{\phantom{a}b} dx^b$.  
Maxwell's equation implies conservation of current $d \star J = 0$ (in 4D). For the toroidally invariant solution this further implies $d * J_P = 0$ (in 2D), which allows to introduce a 0-form current $I$ by writing $* J_P$ as a total derivative 
\ali{
	* J_P = d \left( \frac{I}{\sqrt{-g_T}} \right) \,\, . \label{starJP}
}
The interpretation of $I$ is that it is proportional to the integrated polar current through a `spherical cap' $S$, stretching over a time interval $\Delta t$, azimuthal angle interval  $\Delta \phi = 2\pi$ and range $\theta < \theta_0$ at constant radius $r$, which by Stokes' theorem is given by 
\ali{
	\int_S \star J 
	&= \int_C \sqrt{-g_T} \,  \left(\frac{I}{\sqrt{-g_T}}\right) \, dt \, d\phi   
	= \, (2 \pi  \Delta t) \, I    \label{intpolarcurrent}
} 
with $C$ (along $t$ and $\phi$) the boundary of the 3D surface $S$. Following \cite{Gralla:2014yja}, we will refer to $I$ as the polar current.

We can parametrize the field strength in terms of the polar current $I$. By Maxwell's equations and again Stokes' theorem, the integrated polar current $\int_S \star J$  is also equal to $\int_S d \star F = \int_C \star F = \int_C \star d A_P$, so that $\star d A_P = I \, dt \wedge d\phi$ or 
\ali{
	d A_P = \frac{I}{\sqrt{-g_T}} \sqrt{g_P} \, dr \wedge d\theta 
}
in equation \eqref{F}.  

The FFE equation \eqref{FFEform} with \eqref{starJP} imposes the conditions 
\ali{
	dA_t \wedge dI = 0 \quad \text{and} \quad  dA_\phi \wedge dI = 0 . 
}
These conditions fix different terms of the field strength to be related to each other by 
\ali{
	d A_t = -\, \omega_F \, dA_\phi   \label{dAtofdAphi}
	}
 and 
\ali{
	d I &= f d A_\phi \quad \text{or} \quad I=I(A_\phi) \,.
	}
	 The field strength thus has to be of the form 
\ali{\label{eq:Field}
	F = dA_\phi \wedge \left( d\phi - \omega_F dt \right) + \frac{\sqrt{g_P}}{\sqrt{-g_T}}\, I \, dr \wedge d\theta \,  
}
where $\omega_F$ is the angular velocity of the EM fields, and both $\omega_F$ and the polar current $I$ are functions of $A_\phi$.  

As shown originally in \cite{Uchida2:1997}, plugging this expression for the field strength into the action (\ref{eq:action}) yields an effective 2D action
\begin{align} \label{eq:action}
\mathcal{S}[A] \rightarrow	\mathcal{S}^{eff} [A]	=  -\frac{1}{2} \int \left(   \left|d\phi - \omega_F
	dt \right|^2 |dA_\phi|^2 +\frac{I^2}{g_T} \right) \sqrt{g_P} \sqrt{-g_T} \, dr \,  d\theta \, ,
\end{align}
with the notation $|X|^2 = g^{ab} X_a X_b$ for a 1-form $X$. The integration over $t$ and $\phi$ only produces overall volume factors. This action is a generalization of the Scharlemann-Wagoner action for pulsars \cite{WS} to curved space-times. Similar results  can be obtained from the Einstein-Maxwell action for a field strength of the form (\ref{eq:Field}) with the force-free condition enforced through the use of a Lagrange multiplier term \cite{Thompson:1998ss,Buniy:2013sca}. Caution has to be taken in the process of eliminating $A_P$ from the action using $I(A_\phi)$ \cite{Uchida2:1997}.

The problem is now reduced to solving the remaining FFE equation, also called the stream equation, for the toroidal gauge field $A_{\phi}(x^a)$. It is obtained as the Euler-Lagrange equation of the action  $\mathcal{S}^{eff} [A] = \int \mathcal L \, dr d\theta $, given by    
\begin{equation}
 \partial_{a} \left(\frac{\partial \mathcal{L}}{\partial(\partial_{a}A_{\phi})}\right) -\frac{\partial \mathcal{L}}{\partial A_{\phi}}=0\,. 
\end{equation} 
In the ensuing analysis we will find the action, derive the stream equation and solve this equation for specific choices of the rotating background metrics.

\subsection*{General rotating black hole metric} 
\label{subsec2DPol}

To illustrate the simplicity of this 2D force-free action (\ref{eq:action}), one can take a general stationary and axisymmetric block-diagonal metric
\begin{equation}\label{eq:metric2}
ds^2 = -\alpha^2 dt^2 + \bar \omega^2 (d\phi + \omega \,dt)^2 + (g_{P})_{ab} \, dx^a dx^b \qquad (a,b = r,\theta) \, . 
\end{equation} 
%
Considering a field strength (per construction compatible with the background symmetries) of the form 
\ali{
	\begin{split} 
	F &= dA_\phi \wedge \left(d\phi - \omega_F dt \right) + \frac{\sqrt{g_P}}{\alpha \bar \omega}  \, I\, dr \wedge d\theta \,  
	\end{split}  \label{Fmnansatz}
}
with functions $\omega_F(A_\phi)$ and $I(A_\phi)$, leads to the 2D poloidal system given by  
\ali{
		\mathcal{S}[A] = -\frac{1}{2} \int \sqrt{g_P} \, dr \, d\theta \left( \mathcal C |dA_\phi|^2  - \frac{I(A_\phi)^2 }{\alpha \bar \omega} \right)   \label{S2P}
}
with 
\ali{ 
	\mathcal C =\frac{\alpha}{\bar \omega} - \frac{\bar \omega}{\alpha} (\omega_F + \omega)^2 \,\,.  
}

\section{Near Horizon geometries} 
\label{secNearHorizons}

Rather than analyzing the 2D poloidal problem directly in the Kerr black hole geometry, it has proved advantageous to concentrate specifically on the near-horizon regions, which are described by a decoupled metric. These metrics possess a higher amount of symmetry than the black hole geometry they are derived from through a scaling limit. It is this high amount of symmetry that can be exploited in finding ansatze for symmetric, semi-analytic FFE solutions. This approach has been successfully applied in the near horizon of extremal Kerr or NHEK geometry, in the sense that energy-extracting solutions have been obtained \cite{Lupsasca:2014hua,Lupsasca:2014pfa,Zhang:2014pla}. (However, we will comment on an unphysical feature of such solutions in the discussion of the energy flux.) We generalize the search for energy-extracting FFE solutions to include near-horizon geometries of AdS-Kerr black holes. Indeed, a 3D version of such a solution was presented in \cite{Jacobson:2017xam} for the BTZ black hole. For 
previous work on the BZ process in AdS-Kerr, see \cite{Wang:2014vza}. 

In this section we collect the near horizon geometries that we will consider in this paper, each of the general rotating form \eqref{eq:metric2}: NHEK, AdS-NHEK, SE-NHEK, and near horizon extremal BTZ, which we will refer to as NHEBTZ. 
They are obtained from scaling limits of the respective black hole geometries. These limits are summarized in appendix \ref{appKerr}. The action and stream equation associated to these specific near horizon geometries are also described.

\subsection{Near Horizon Extreme geometries}  

In this paper we are interested in the region very near the horizon of extreme Kerr and AdS-Kerr. These regions are described
by the so-called Near-Horizon Extreme geometries defined by
\begin{eqnarray}\label{NHE}
ds^2=\Gamma(\theta) \left[-r^2 dt^2+\frac{dr^2}{r^2}+\alpha(\theta)^2d\theta^2+\gamma(\theta)^2(d\phi+k\, r \,dt)^2\right]\,,
\end{eqnarray}
and characterized in each case by specific functions $\Gamma(\theta), \alpha(\theta),\gamma(\theta)$ and parameter $k$. A separate discussion of each geometry is given in the next subsections.

\subsubsection{Near Horizon Extreme Kerr}  
\label{sectionNHEK}

The line element of the so-called Near-Horizon Extreme Kerr (NHEK) geometry \cite{Bardeen:1999px} is of the form (\ref{NHE}) with
\ali{
	  \Gamma(\theta) = M^2 \left(1 + \cos^2 \theta \right) \, ,\qquad\alpha(\theta)^2=1\,, \qquad \gamma(\theta) =\Lambda(\theta)= \frac{2 \sin \theta}{1 + \cos^2 \theta}  \,,\qquad k=1 \label{ds2NHEK} 
} 
and metric determinant $\sqrt{-g}= \Lambda\,\Gamma^2$. It is the decoupled near horizon limit of extremal Kerr and co-rotates with the extremal Kerr metric at the angular velocity of the horizon $\Omega_H^{ext} = \frac{1}{2M}$, see equation \eqref{OmegaHextNHEK}. 

In contrast with the original Kerr metric, the NHEK geometry is not asymptotically flat; the metric contains an AdS$_2$ factor in the $(r,t)$ directions and correspondingly an  $\SL(2,\R)$ isometry. The full isometry group is $\SL(2,\R)\times\U(1)$.  The $\U(1)$ rotational symmetry is generated by the Killing vector field 
\begin{align}
\label{W0NHEK}
W_0=\pd_\phi 
\end{align}
and the $\SL(2,\R)$ symmetry by the Killing vector fields 
\begin{align}
H_0&=t\pd_t-r\pd_r \, ,  \label{HNHEK} \\ 
H_+&=\sqrt{2}\pd_{t} \, ,\label{eq:H+NHEK}\\
H_-&=\sqrt{2}\br{\frac{1}{2}\pa{t^2+\frac{1}{r^2}}\pd_t-t \,  r\pd_r-\frac{1}{r}\pd_{\phi}} \, .\label{eq:H-NHEK}
\end{align}
It is easily verified that these  satisfy the $\SL(2,\R)\times\U(1)$ commutation relations 
\begin{align} 
\begin{split} 
\br{H_0,H_\pm}&=\mp H_\pm \, ,\qquad\,\br{H_+,H_-}=2H_0 \, ,\\
\br{W_0,H_\pm}&=0 \, ,\qquad\qquad\br{W_0,H_0}=0 \, . 
\end{split}  \label{commrels}
\end{align} 

\subsubsection{Near Horizon Extreme AdS-Kerr}
\label{sectionAdSNHEK} 

The near horizon of extreme AdS-Kerr (AdS-NHEK) geometry \cite{Hartman:2008pb} is given by (\ref{NHE}) where
\begin{eqnarray}\label{eq:AdsDefs}
\Gamma(\theta)=\frac{r_+^2 +a^2\cos^2\theta}{\Delta_0}\,,\qquad\alpha(\theta)^2=\frac{\Delta_0}{1-(a^2/l^2)\cos^2\theta}\,,\qquad \gamma(\theta)=\frac{(r_+^2+a^2)\sin\theta}{\alpha(\theta)\,\Gamma(\theta)\,\Xi}\,,\, 
\end{eqnarray}
and constants $\Delta_0=1+a^2/l^2+6 r_+^2/l^2$, $\Xi=1-a^2/l^2$ and $k= 2 \,a \,r_+\, \Xi/(\Delta_0 (r_+^2 +a^2))$. This metric obeys $R_{\mu\nu}=-3\,l^{-2} g_{\mu\nu}$. 
It is of the form ($\ref{eq:metric2}$) with 
$\sqrt{-g_T}=r\,\gamma\,\Gamma$,   $\sqrt{g_P}=\alpha\,\Gamma/r$ and $\sqrt{-g}=\alpha\,\gamma\,\Gamma^2$. 

The parameter $r_+$ is defined as the largest root of 
\begin{eqnarray}\label{eq:horizon}
(l^2+a^2+3 \,r_+^2) \, r_+^2-l^2 a^2=0\,.
\end{eqnarray}
Finally note that the NHEK geometry (\ref{ds2NHEK}) is recovered in the limit $l\rightarrow\infty$ while  %
\begin{eqnarray}
k\rightarrow 1\,,\qquad \alpha(\theta)\rightarrow 1\,,\qquad \gamma(\theta)\rightarrow \Lambda(\theta)  \,.
\end{eqnarray}

As in NHEK, a crucial feature of the AdS-NHEK region is that the original  $\U(1)\times\U(1)$ Kerr-AdS isometry group is enhanced to $\SL(2,\R)\times\U(1)$. In fact, AdS-NHEK has the same isometries as NHEK up to a rescaling of $\phi$. The $\U(1)$ rotational symmetry is generated by the Killing vector field $W_0$ in \eqref{W0NHEK}. The time translation symmetry becomes part of an enhanced $\SL(2,\R)$ isometry group generated by the Killing vector fields $H_0$ in \eqref{HNHEK}, $H_+$ in \eqref{eq:H+NHEK} and  
\begin{align} 
H_-&=\sqrt{2}\br{\frac{1}{2}\pa{t^2+\frac{1}{r^2}}\pd_t-t \, r\pd_r-\frac{k}{r}\pd_{\phi}} \, .\label{eq:H-} 
\end{align} 

The Killing vectors satisfy the $\SL(2,\R)\times\U(1)$ commutation relations in \eqref{commrels}. 

\subsubsection{Super-Entropic Near Horizon Extremal AdS-Kerr}
\label{sectionSENHEK} 

The near horizon geometry of the super-entropic extremal AdS-Kerr black hole \cite{Sinamuli:2015drn}, or simply SE-NHEK metric, is of the form $(\ref{NHE})$   where now
\begin{eqnarray}\label{eq:SEdefd}
{\Gamma}(\theta)=\frac{l^2(1+3\cos^2\theta)}{12}\,,\qquad {\alpha}(\theta)^2=\frac{4}{\sin^2\theta}\,,\qquad {\gamma}(\theta)=\frac{4\,l^2}{3}\frac{\sin\theta}{{\alpha}(\theta)\,{\Gamma}(\theta)}\,,\qquad {k}=\frac{\sqrt{3}}{8}\,.\quad   
\end{eqnarray}
There are essentially two ways to derive this metric. In Appendix \ref{subsec:SENEHK} we describe how to obtain the near horizon geometry from the extremal super-entropic AdS-Kerr black hole \cite{Cvetic:2010jb,Hennigar:2015cja}. A more straightforward way is to start with the AdS-NHEK line element (\ref{NHE}) with (\ref{eq:AdsDefs}), and before taking $a\rightarrow l$ rescale ${\phi} \ra {\phi} \,\Xi$. The coordinate can be chosen such that $\phi \sim \phi +2\pi$ (for the new angle $\phi$).  This geometry has the special feature that the locations $\theta=0,\pi$ are removed from the space-time. The isometry group of the SE-NHEK geometry is $\SL(2,\R)\times\U(1)$ as for AdS-NHEK, with the same Killing vector fields \eqref{W0NHEK}-\eqref{eq:H+NHEK} and $(\ref{eq:H-})$ (with $k$ in \eqref{eq:SEdefd}).

\subsubsection{Near Horizon Extreme BTZ}
\label{sectionNHBTZ} 
For future reference we also describe the near horizon geometry of the extreme BTZ (NHEBTZ) black hole metric \cite{Coussaert:1994tu,Balasubramanian:2009bg,deBoer:2010ac}. 
The line-element is  
\ali{
	ds^2 
	&= \frac{l^2}{4} \frac{dr^2}{r^2} + 2 \frac{r}{l} dt d\phi  + r_+^2 d\phi^2 . \label{ds2nhBTZ} 
} 
The geometry retains all the relevant aspects of black holes, i.e.~a horizon, and necessary ingredients for energy extraction, i.e.~an ergosphere. While the NHEBTZ region provides a rich context to study FFE, the focus of our work will remain primarily in 4 dimensions.  On the BTZ spacetime,  purely electromagnetic versions of the BZ process were found, in which plasma surprisingly plays no role \cite{Jacobson:2017xam}. 
In section \ref{subsec:BTZ}, we employ these exact analytical FFE solutions in BTZ to argue that our prescription reproduces the energy flux from the extreme black hole throat.

\subsection{Action and stream equation} 
\label{subsecActions}

We now turn to the study of the force-free electrodynamics action in the near horizon geometries. For a near horizon extremal metric \eqref{NHE}, the action (\ref{eq:action}) is  
\ali{
	\begin{split} 
		\mathcal{S}[A_\phi] &= \int dr d\theta \left[ \frac{d_\alpha}{2} \left(r^2 (\p_r A_\phi)^2 + \frac{1}{\alpha^2} (\p_\theta A_\phi)^2 \right) +  \frac{\alpha}{2 r^2 \gamma} I(A_\phi)^2 \right]    \\
		&\qquad \qquad  d_\alpha \equiv -\frac{\alpha}{r}\mathcal C = \alpha \, \gamma \left( \frac{\omega_F}{r}+k \right)^2 -\frac{\alpha}{\gamma}  \, .
	\end{split} \label{generalAction}
}
The relevance of $\mathcal C$ is that it becomes zero at the so-called light surface or light cylinder. At these locations, an observer co-rotating with the field lines would have to travel at the speed of light. 

Variation of this action gives the  
stream equation for $A_\phi$, 
\ali{
	\p_\theta ( d_\alpha \, \frac{1}{\alpha^2} \p_\theta A_\phi) + \p_r ( d_\alpha \, r^2 \p_r A_\phi) - \frac{1}{2} \p_{A_\phi} d_\alpha 
	\left( r^2 (\p_r A_\phi)^2 + \frac{1}{\alpha^2}(\p_\theta A_\phi)^2 \right) - \frac{ \alpha}{r^2 \gamma} I \, \p_{A_\phi} I  
	= 0 \, .   \label{AEOM}
}
Further details about the actions and stream equations can be found in Appendix \ref{appEOM}. Observe that the stream equation (\ref{AEOM}) is highly nonlinear and can in general only be solved numerically. Moreover, the possibility of $\mathcal C$ having zeros in the domain of integration may lead to further complications. Despite these issues, the symmetries can be exploited to simplify the analysis. We will show that imposing scaling symmetry allows to solve the force-free equations semi-analytically and find energy extracting configurations. 


\section{Force-Free Electrodynamics solutions} \label{sectionFFEsols} 

In this section we construct solutions to the FFE stream equations in (AdS/SE-)NHEK. The stream equation \eqref{AEOM} for the gauge field component $A_\phi(r,\theta)$ is written out explicitly for the (AdS/SE-)NHEK background geometries (see also Appendix \ref{appEOM}). To obtain the stream equations in the form  \eqref{AEOM} we already have used one aspect of the symmetry of the problem, that is, the gauge field solution is required to have the same independence on time $t$ and azimuthal angle $\phi$ as the stationary and axisymmetric background metrics. 
This strategy can be applied further by also imposing the scaling symmetry of the background near horizon metrics on the solution. This gives rise to two ansatze for the behavior of the solution as a function of the radius $r$, which we will refer to as the `scaling ansatz' (which was considered earlier in \cite{Zhang:2014pla,Lupsasca:2014hua}) and the `log ansatz'. For these ansatze, the stream equations reduce further to a single ODE for the $\theta$-profile of $A_\phi$, which can be solved numerically. The ODE's are summarized in section \ref{sectionEOM}.

\subsection{Ansatze} 

Consider the stationary and axisymmetric field strength ansatz \eqref{Fmnansatz}, written out explicitly to 
\ali{
	F = \p_r A_\phi dr \wedge d\phi +  \p_\theta A_\phi d\theta \wedge d\phi - \omega_F \p_r A_\phi dr \wedge dt - \omega_F \p_\theta A_\phi d\theta \wedge dt  + \frac{I}{\alpha \bar \omega} \sqrt{g_P} dr \wedge d\theta  \label{Fansatz}
}
where $\frac{\sqrt{g_P}}{\alpha \bar \omega}$ scales as $1/r^2$ for each of the near horizon geometries we consider. In the form \eqref{Fansatz}, it is easy to read off the conditions for self-similar behavior of the field strength under scaling transformations.      

\subsubsection{Scaling ansatz}  
\label{sectionscaling}

The field strength \eqref{Fansatz} will display scaling symmetry 
\ali{
	F \ra \lambda^h F 
}
under a scaling transformation $r \ra \lambda \, r$, $t \ra t/\lambda$ generated by $H_0 = r \p_r - t \p_t$ in \eqref{HNHEK}, only if  
\ali{
	\p_r A_\phi \sim r^{h-1} , \quad \p_\theta A_\phi \sim r^h , \quad \omega_F  \sim r ,\quad I \sim r^{h+1}\, . 
}
These conditions are solved by the `scaling ansatz' for $A_\phi$ in terms of a field $\Phi(\theta)$, and $I$ and $\omega_F$ as functions of $A_\phi$:   
\ali{
	\begin{split} 
	A_\phi &= r^h \Phi(\theta), \qquad h \neq 0 \\ 
	I &= I_0 \, A_\phi^{1+1/h} \\
	\omega_F &= \omega_0 A_\phi^{1/h} 
	\end{split} \label{scalingansatz}
}
where $I_0$ and $\omega_0$ are integration constants. By construction, the scaling ansatz field strength satisfies the symmetries of the (AdS/SE-)NHEK region  
\ali{
	\mathcal{L}_{W_0}F=0\,,\qquad\,\mathcal{L}_{H_+}F=0\,,\qquad\mathcal{L}_{H_0}F=-h F\, \, .   \label{Finvariance} 
}
Moreover, the scaling ansatz vector potential,
with $A_t = \omega_0 A_\phi^{1+1/h}/(1+1/h) + c$ satisfies the same symmetries, 	$\mathcal{L}_{W_0}A=0$, $\mathcal{L}_{H_+}A=0$ and $\mathcal{L}_{H_0}A=-h A$,  if the constant $c$ equals zero. 

\subsubsection{Log ansatz}   \label{sectionlogansatz}

The $h=0$ case needs to be handled separately. Imposing the scaling symmetry 
\ali{
	F \ra F 
}
under the scaling transformation $r \ra \lambda \, r$, $t \ra t/\lambda$, requires 
\ali{
	\p_r A_\phi \sim r^{-1} , \quad \p_\theta A_\phi \sim 1 , \quad \omega_F  \sim r ,\quad I \sim r.  
}
This leads to the `log ansatz' in terms of a field $\Phi(\theta)$: 
\ali{
	\begin{split} 
	A_\phi &= \log (r \, \Phi(\theta))/\omega_0 \\
	I &= I_0 \,  r \, \Phi(\theta)  \\
	\omega_F &= \omega_0 \,  r \, \Phi(\theta)  
	\end{split} \label{logansatz}
}
with constants $I_0$ and $\omega_0$. The same invariance \eqref{Finvariance} of the field strength applies to this log ansatz. However, the gauge field is not in a highest-weight representation of $\SL(2,\R)$ as  $\mathcal{L}_{H_0}A \neq 0$. In this sense the log ansatz could be called an approximate symmetric ansatz.

\subsection{Equations of motion} \label{sectionEOM} 

First let us focus on the NHEK background. 
The stream equation \eqref{AphiEOMNHEK} for a field $A_\phi$ of the form
specified in the scaling ansatz \eqref{scalingansatz} reduces to an ODE for the field $\Phi(\theta)$, which we can call the `NHEK scaling EOM': 
\ali{
	\begin{split} 
	&\p_\theta \left(d \, \p_\theta \Phi \right) + d \, h (h+1) \Phi - \frac{1}{2} (\p_\Phi d) \left(h^2 \Phi^2 + (\p_\theta \Phi)^2 \right) - \frac{1}{\Lambda} I_0^2 (1 + \frac{1}{h}) \Phi^{1+2/h} = 0 \\ 
	&\qquad \qquad \text{with} \quad d = \Lambda ( \omega_0 \Phi^{1/h} + 1)^2 - \frac{1}{\Lambda} \, .    \label{NHEKscalingEOM}
	\end{split} 	
}
Here $d$ is itself a functional of $\Phi(\theta)$ and is defined in \eqref{generalAction} to be proportional to the functional $\mathcal C$, which determines the location of light surfaces.  

Evaluating the EOM \eqref{generalAction} for the log ansatz \eqref{logansatz} leads to 
the `NHEK log EOM' for $\Phi(\theta)$, 
\ali{
	\begin{split} 
	&\p_\theta \left(D \, \p_\theta \Phi \right) - \frac{1}{2} (\p_\Phi D) \left(\Phi^2 + (\p_\theta \Phi)^2 \right) -  \frac{1}{\Lambda} I_0^2 \Phi = 0   \\
	&\qquad\qquad \text{with} \quad D = \Lambda \left(1 + \frac{1}{\omega_0 \Phi}\right)^2 -\frac{1}{\Lambda} \frac{1}{\omega_0^2 \Phi^2} \, . 
	\end{split} 
}
Similarly, for a near-horizon region of AdS-Kerr (\ref{NHE}), the `AdS-NHEK scaling EOM' 
\ali{
	&\p_\theta \left(d_\alpha \frac{1}{\alpha^2} \p_\theta \Phi \right) + d_\alpha \, h (h+1) \Phi - \frac{1}{2} (\p_\Phi d_\alpha) \left(h^2 \Phi^2 + \frac{1}{\alpha^2}(\p_\theta \Phi)^2 \right) - \frac{ \alpha}{\gamma} I_0^2 (1 + \frac{1}{h}) \Phi^{1+2/h} = 0  \nonumber \\
	&\qquad \qquad \qquad \text{with} \quad d_\alpha = \alpha \gamma ( \omega_0 \Phi^{1/h} - k)^2 + \frac{\alpha}{\gamma} 
	\label{AdSNHEKscalingEOM}
}
is the EOM \eqref{AphiEOMAdS} for the scaling ansatz \eqref{scalingansatz}. The metric functions $\alpha, \gamma$ and $k$ are specified in  \eqref{eq:AdsDefs} for AdS-NHEK and in \eqref{eq:SEdefd} for SE-NHEK. 
The `AdS-NHEK log EOM'  
\ali{
	\begin{split} 
	&\p_\theta \left(D_\alpha \frac{1}{\alpha^2} \, \p_\theta \Phi \right) - \frac{1}{2} (\p_\Phi D_\alpha) \left(\Phi^2 + \frac{1}{\alpha^2}(\p_\theta \Phi)^2 \right) - \frac{ \alpha}{\gamma} I_0^2 \Phi = 0   \\ 
	&\qquad \qquad \text{with} \quad D_\alpha = \alpha \gamma \left(1 +\frac{k}{\omega_0 \Phi}\right)^2 -\frac{\alpha}{\gamma} \frac{1}{\omega_0^2 \Phi^2} 
	\end{split}  \label{SENHEKlogEOM}
}
is the EOM \eqref{AtEOMAdS} for the log ansatz \eqref{logansatz}. 

The functionals $d, D, d_\alpha$ and $D_\alpha$ introduced here  become zero at the location of a light surface. For the solutions that we will discuss, these functionals will have a definite sign.  

\subsection{Boundary conditions} \label{subsec:bc}

When solving the EOM's in the previous section, the physical boundary conditions we impose on the field $\Phi(\theta)$ are the same as in \cite{Zhang:2014pla}. 
The range of the angle $\theta$ stretches from $\theta = 0$ at the north pole, over $\theta = \pi/2$ at the equator, to $\theta = \pi$ at the south pole.  
For the solution to be north-south symmetric, i.e.~symmetric under reflection $\theta \ra \pi - \theta$, we require the field to have a vanishing derivative at the equator, 
\ali{ \Phi'(\pi/2) = 0 .  \label{bc1}}    
A second condition follows from considering the integrated polar current through a spherical cap $S: \theta < \theta_0$, as given in equation \eqref{intpolarcurrent}, $\int_{C = \p S} \star F = \, (2 \pi \Delta t) \, I$. 
Consider the limit that the spherical cap shrinks to the point at the north pole $\theta_0 \ra 0$. For a physical solution with non-diverging field strength $F$ at the north pole,  the integrated polar current will vanish in this limit.  For both our ansatze, \eqref{scalingansatz} and \eqref{logansatz}, the vanishing of the polar current $I$ at $\theta = 0$ imposes the field $\Phi$ to vanish in that point. The second boundary condition is therefore \ali{ 
	\Phi(0) = 0.  \label{bc2}
}

\subsection{Solutions} \label{sectionsols} 

A shooting method is employed to numerically solve the EOM's of section \ref{sectionEOM} using NDSolve in Mathematica.  
We recover a NHEK scaling solution that has previously been discussed in the literature, but have not succeeded in finding a NHEK log solution that satisfies the boundary condition \eqref{bc2}. The new solutions we find are an AdS-NHEK scaling solution and an SE-NHEK log solution. The $\Phi(\theta)$ profiles for the numerical solutions are presented in Figures  \ref{figscalingNHEK}-\ref{figSENHEKlogII}. To determine whether they are energy-extracting, we turn now to a discussion of the energy flux.  \\

In the following sections, we will evaluate the energy and angular momentum flux densities of near horizon geometries to show that our force-free solutions do indeed produce non-trivial fluxes measured by 
stationary observers outside the near horizon throats. 


\section{Energy flux from extreme black hole throat} 
\label{sectionenergy}

Before analyzing in this section the energy flux of our obtained solutions in rotating geometries, we first turn to a toy model. The toy model consists of a rotating electromagnetic (EM) field configuration, and serves to illustrate that the direction of the Poynting flux, i.e.~whether the set-up has an inwards or outwards pointing flux of energy,  depends on the rotation of the observer. 
The lesson relevant for the discussion of the energy flux of the (AdS/SE-)NHEK solutions, is that it is the Poynting flux measured by a Kerr observer rather than a near horizon observer (which rotate with respect to each other) that determines whether or not a solution is energy extracting.

\subsection{Toy model}  \label{toymodelsection}

\begin{figure}
	\includegraphics[width=8.cm]{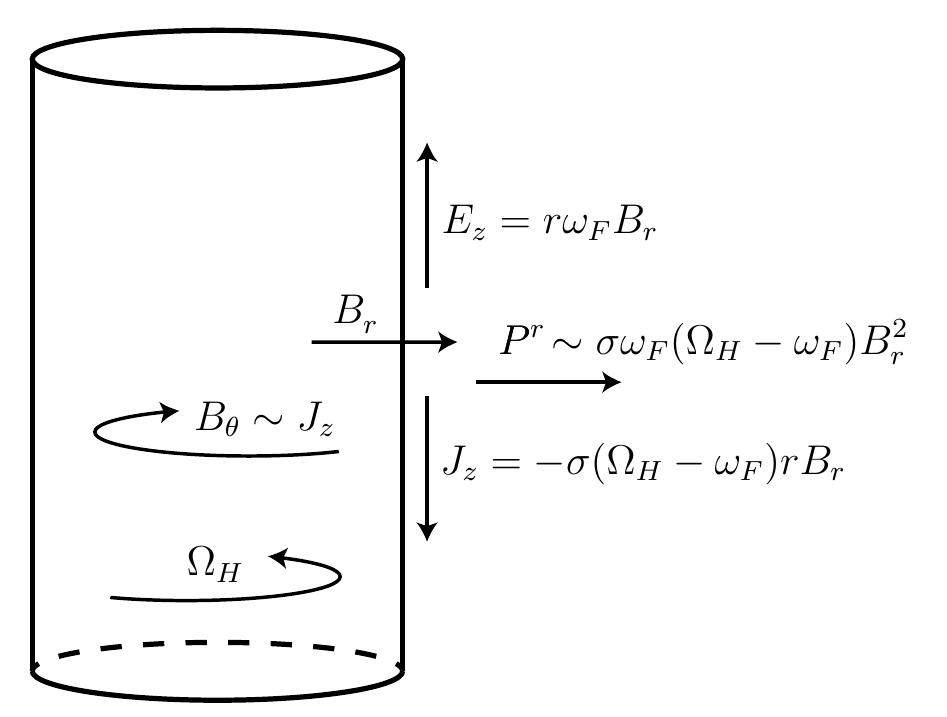} \quad \includegraphics[width=8.cm]{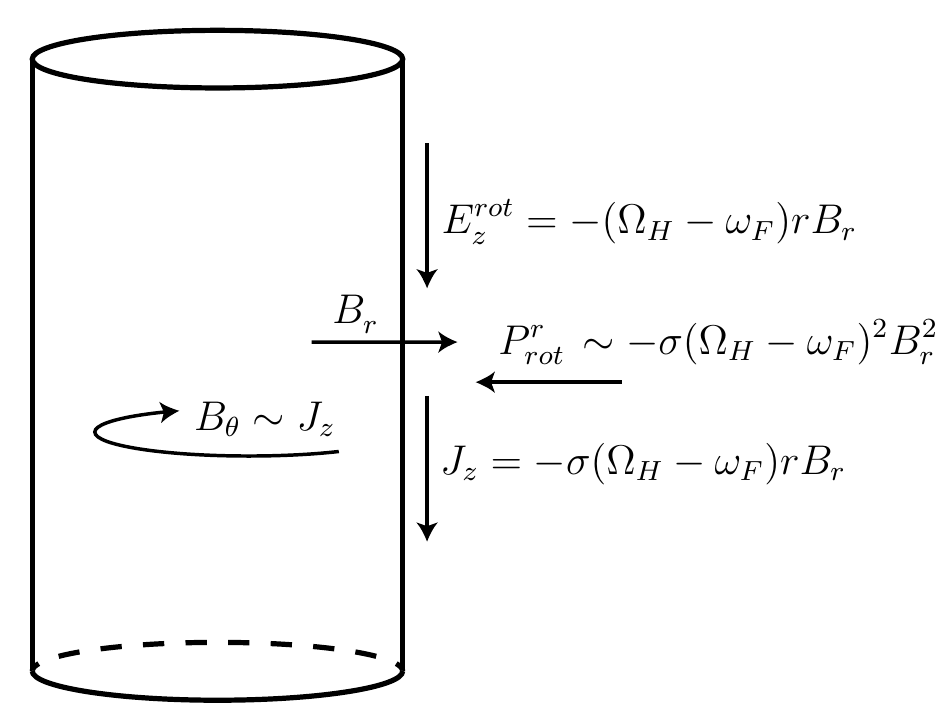} \\
	\caption{Toy model set-up of a rotating conducting cylinder in the lab frame (left) and a frame that rotates with the cylinder (right).}  \label{figtoymodel}
\end{figure}

Consider a conducting cylinder that rotates with an angular velocity $\Omega_H$ and is placed in a radial magnetic field $B_r$. This is an academic set-up that we consider for the sake of argument of describing the energy outflux in a rotating frame.      	
The set-up has non-zero fields 
\ali{ 
	E_z = r \omega_F B_r(r), \quad B_\theta = B_\theta(r) , \quad  B_r = B_r(r).  \label{Libbytoysetup}
}
The $E$-field contributes a term $\sigma \omega_F r B_r$ proportional to the conductivity $\sigma$ to the current $J_z$. Because the conducting cylinder moves with a velocity $v$, the charges in it feel an electromotive force $q v \times B$, adding another term to the current. 
In total, $J_z = \sigma (\omega_F - \Omega_H) r B_r$. The current induces, by the integral Ampere law, the $B$-field $B_\theta \sim J_z \sim \sigma (\omega_F - \Omega_H) r^2 B_r$. 

The Poynting vector, as defined in \eqref{Plabcyl}, is $P = E \times B = (-\omega_F B_r B_\theta,\omega_F B_r^2,0)$ for $P = (P^r,P^\theta,P^z)$. Its radial component is 
\ali{
	P^r = - \frac{B_\theta E_z}{r} = -\omega_F B_r B_\theta \sim  \sigma r^2 B_r^2 \omega_F (\Omega_H - \omega_F)  \label{Poyntinglab}
} 
so that the sign of the vector is determined by the sign of  $\Omega_H - \omega_F$. 

The system can also be observed from a rotating frame with angular velocity $\omega = \Omega_H$ or $\theta' = \theta - \Omega_H t$. In the rotating frame, the contribution to the current that in the lab frame originated from an electromotive force, has to follow from an electric field contribution. The total observed electric field (in the sense that it obeys
Faraday's equation in the rotating frame, see \eqref{Erot} in Appendix \ref{appEM}) is given by $E_z^{rot} = r (\omega_F - \Omega_H) B_r$. 
The observed magnetic fields \eqref{Brot} remain the same as in \eqref{Libbytoysetup}.  
The Poynting flux in the rotating frame\footnote{Note that the mixing of $E$ and $B$ fields in rotating frames makes the definitions of $E$ and $B$ ambiguous, reflected in the Maxwell equations \eqref{rotatingmaxwell}-\eqref{arotatingmaxwelleq} for $E,B,E_{rot},B_{rot}$. We refer to Appendix \ref{appEM} for a discussion of EM fields in a rotating frame.}  can then be defined as $P_{rot} = E_{rot} \times B$. It has radial component 
\ali{ 
	P^{r}_{rot} = - \frac{B_\theta E_z^{rot}}{r} = -B_\theta (\omega_F - \Omega_H) B_r \sim - (\omega_F - \Omega_H)^2 r^2 B_r^2. \label{Poyntingrot}
} 
This is the Poynting flux in the right figure of Figure \ref{figtoymodel}. It is inwards pointing, even though the physical Poynting flux in the lab frame \eqref{Poyntinglab} is outwards pointing. \\

Let us repeat the analysis of the toy model in a covariant language that will be straightforwardly applicable to the force-free electrodynamics in a rotating NHEK frame versus  asymptotically non-rotating Kerr frame. 

The field strength in the lab frame \eqref{ds2cyl} is 
\ali{
	F_\mn^{lab} = B_r r \left(d\theta - \omega_F dt \right) \wedge dz - \frac{B_\theta}{r} dr \wedge dz 
}
and the field strength in a rotating frame \eqref{rotcylmetric} (with $\theta' = \theta - \omega t$) is 
\ali{ 
	F_\mn^{rot} = B_r r \left( d\theta' + (\omega - \omega_F) dt \right) \wedge dz - \frac{B_\theta}{r} dr \wedge dz \, . 
} 
In particular, we will consider a frame that co-rotates  with the cylinder, $\omega = \Omega_H$.  

The conserved energy flux seen by a stationary observer in the lab frame can be calculated from the Maxwell stress-energy tensor 
\ali{
	T_\mn = -\frac{1}{4} g_\mn F_{\alpha\beta}F^{\alpha \beta} + F_{\mu \alpha} F_\nu^{\phantom{\nu}\alpha}  \label{stresstensor}
}
as 
\ali{
	\mathcal E^\mu = - T^\mn \chi_\nu  \label{energyflux}
}
where $\chi=\p_t$ is the global timelike Killing vector. Similarly, the angular momentum flux   
\ali{
	\mathcal L^\mu = T^\mn \eta_\nu  \label{Lflux}
}
is defined in terms of the axial Killing vector $\eta = \p_\theta$.  Here and hereafter, 
it is understood that we have dropped the `$EM$' 
label in the symbol $T^{EM}_\mn$ to denote the electromagnetic stress-energy tensor defined in \eqref{stresstensor}. 
The transformation matrix for the coordinate transformation from the lab frame, in cylindrical coordinates $x^\mu = (t,r,\theta,z)$, to the rotating frame $x^{\mu'} = (t',r',\theta',z') = (t,r,\theta-\Omega_H t,z)$ is given by 
\ali{
	\frac{\p x}{\p x'}  = \left( \begin{array}{cccc} 1&0&0&0 \\ 0&1&0&0 \\ \Omega_H&0&1&0 \\ 0&0&0&1  
	\end{array} \right) .   \label{Lambdarotatingcyl}
}
Under the coordinate transformation, the energy flux transforms to 
\ali{
	\mathcal E^{\mu'} =  
	\frac{\p x^{\mu'}}{\p x^{\mu}} \mathcal E^{\mu}.   \label{toymodel48}
}
In particular, $\mathcal E^{r'} = \mathcal E^{r}$. This is the lab energy flux as measured by a rotating observer. 
In terms of the stress tensor in the rotating frame, it is given by 
\ali{
	\mathcal E^{\mu'}  = - T^{\mu'}_{\phantom{\mu}\nu'} \chi^{\nu'} = -(T^{\mu'}_{\phantom{\mu'}t'} - \Omega_H T^{\mu'}_{\phantom{\mu'}\theta'})   \label{toymodel49}
}
where we made use of $\chi^{\nu'} = \frac{\p x^{\nu'}}{\p x^{\mu}}  \chi^\mu = \frac{\p x^{\nu'}}{\p t} = (1,0,-\Omega_H,0)$.  
It follows that the radial energy flux in the lab frame is obtained from a combination of the radial energy flux and angular momentum flux in the rotating frame 
\ali{
	\mathcal E^{r} = \mathcal E^{r'}_{rot} + \Omega_H \mathcal L^{r'}_{rot} \,\, ,  \label{flux}
}
where we introduced the notation $\mathcal E^{r'}_{rot} \equiv  -T^{r'}_{\phantom{\mu'}t'}$ and $\mathcal L^{r'}_{rot} \equiv T^{r'}_{\phantom{\mu'}\theta'}$ in analogy with \eqref{energyflux}-\eqref{Lflux}. 
We would like to point out that there is no ambiguity of interpretation in equation \eqref{flux}. It relates a component of the lab flux to components of the rotating frame fluxes, by making use of $\mathcal E^{\mu} = -\frac{\p x^{\mu}}{\p x^{\mu'}} (T^{\mu'}_{\phantom{\mu'}t'} - \Omega_H T^{\mu'}_{\phantom{\mu'}\theta'} )$. In this simple set-up where $r'=r$, we can change $r'$ to $r$ on the right hand side of \eqref{flux}.

From equation \eqref{ProtfromT} in the general discussion in the appendix, the energy flux $\mathcal E^{r'}_{rot} = r(\Omega_H-\omega_F) B_r B_\theta$ equals the Poynting flux $P^{r}_{rot}$ in \eqref{Poyntingrot}, and $\mathcal L^{r}_{rot} = -r B_r B_\theta$.  
The relation \eqref{flux} thus relates the radially inwards Poynting vector measured in the rotating frame 
to the outwards one measured in the lab frame   
\ali{
	P^{r} = P^{r}_{rot} + \Omega_H \mathcal L^{r}_{rot} \,\,,   
}
consistent with equations \eqref{Poyntinglab} and \eqref{Poyntingrot}. The outflux $P^r$ in \eqref{Poyntinglab} is positive when $\Omega_H > \omega_F$, and rotational energy can be transferred from the cylinder to the field. The suggestive notation $\Omega_H$ for the angular velocity of the cylinder indicates that we will  
think of the rotation of a black hole in an analogous way.

\subsection{Energy flux in Kerr} 

In this section we derive the analogue of the relation \eqref{flux} 
for the energy flux measured in a Kerr geometry as a function of the flux measured by an observer in the near-horizon region. 

The NHEK geometry \eqref{ds2NHEK}, with $x^\mu=(t,\phi,r,\theta)$,   
is obtained from the Kerr geometry in Boyer-Lindquist coordinates $x^{\hat \mu} =(\hat t,\hat \phi,\hat r,\hat \theta)$ through the coordinate transformation (see \eqref{KerrtoNHEK}) 
\ali{
	\hat t = \frac{t}{\zeta} , \qquad \hat \phi = \phi + \frac{t}{2M\zeta}, \qquad \hat r = 2M^2 \zeta r + M , \qquad \hat \theta = \theta  \label{coordtransf}
}
in the limit where the scaling parameter $\zeta$ vanishes. 
The transformation matrix is 
\ali{
	\frac{\p \hat x}{\p x} = \left( \begin{array}{cccc} \frac{1}{\zeta}&0&0&0 \\ \frac{1}{2M\zeta}&1&0&0 \\ 0&0& 2M^2 \zeta &0 \\ 0&0&0&1  
	\end{array} \right) \,\, . 
}

We are interested in the energy flux seen by a stationary Kerr observer (with hatted coordinates). It is defined in terms of the Kerr stress tensor $T^{\hat \mu \hat \nu}$ and the Kerr timelike Killing vector $\chi^{\hat \mu} = (1,0,0,0)$ as \cite{Blandford:1977ds} 
\ali{ 
	\mathcal E^{\hat \mu}
	\equiv -T^{\hat \mu\hat \nu} \chi_{\hat \nu} = -T^{\hat \mu}_{\phantom{\mu }\hat t} \,\, .  \label{Ehatdef}
}  
The solutions we obtained are solutions in the NHEK geometry (with unhatted coordinates). We therefore need to work out first what the relation is between $\mathcal E^{\hat \mu}$ and quantities measured in NHEK. 
This follows closely the discussion in section \ref{toymodelsection} of the toy model (equations \eqref{toymodel48} and \eqref{toymodel49}), with the Kerr frame taking the role of the lab frame, and the NHEK frame the role of the rotating frame.  

If we introduce the notation\footnote{Note that $\mathcal E^r
	$ is not the radial energy flux observed by a stationary NHEK observer, which would be defined as $-T^{\mn} \xi_{\nu} = -T^\mu_{\phantom{\mu}t}$ in terms of the NHEK timelike Killing vector $\xi$. When it is necessary to distinguish this object, we will refer to it as $\mathcal E^\mu_{rot} \equiv -T^\mu_{\phantom{\mu}t}$, following the notation in equation \eqref{flux} in the toy model.}
\ali{
	\mathcal E^\mu
	\equiv -T^{\mn} \chi_{\nu} 
}
in terms of the transformed $T^{\mn}$ and $\chi^\mu$, then we find 
\ali{
	\mathcal E^r
	= \frac{1}{2M^2 \zeta} \mathcal E^{\hat r}.  
} 
This gives the scaling between the energy flux rate per NHEK time, $\mathcal E^r$, and the energy flux rate per Kerr time, $\mathcal E^{\hat r}$, and expresses that the scale invariant quantity is the energy $\mathcal E^r \Delta t \sim \mathcal E^{\hat r} \Delta \hat t$. 

Furthermore, because the stationary Kerr observer is rotating from the NHEK-perspective 
\ali{
	\chi^{\mu} = \frac{\p x^\mu}{\p x^{\hat \mu}} \chi^{\hat \mu} = \frac{\p x^\mu}{\p \hat t} = \left( \zeta,-\frac{1}{2M},0,0 \right),   \label{rotNHEKobs}
}
it follows that 
\ali{ 
	\mathcal E^r = -\left(\zeta T^r_{\ph{\mu} t} - \frac{1}{2M} T^r_{\ph{\mu} \phi} \right) .   \label{ErNHEK}
} 
We thus find the following expression for the Kerr energy flux as a function of NHEK observables 
\ali{
	\mathcal E^{\hat r}	= -\frac{\lambda^2}{2}  \left( T^r_{\ph{\mu} t} - \frac{1}{\lambda} T^r_{\ph{\mu} \phi} \right) ,   \label{ErKerr}
}
which follows most directly from the transformation $T^{\hat \mu}_{\phantom{\mu} \hat \nu} = T^\mu_{\phantom{\mu}\nu} \frac{\p x^{\hat \mu}}{\p x^\mu} \frac{\p x^\nu}{\p x^{\hat \nu}}$.  
Here the scaling parameter 
\ali{
	\lambda = 2 M \zeta   \label{lambdaNHEK}
}
was introduced, which has the following interpretation. 
By writing the Kerr to NHEK angle transformation as 
\ali{
	\hat \phi = \phi + \frac{1}{\lambda} t \,\, , 
}
the scale factor can be interpreted as the angular velocity $\Omega_A^{NHEK}$ of the asymptotic (Kerr) region as seen by a NHEK observer 
\ali{
	\Omega_A^{NHEK} = \frac{1}{\lambda}. 
}   
The 
rotation $\omega = r$ of the NHEK metric (of the form \eqref{eq:metric2}) should not exceed the value $\Omega_A^{NHEK}$, and therefore the `gluing' from NHEK to the asymptotic Kerr region should take place at 
\ali{
	r_* = \frac{1}{\lambda}. 
}
This makes the statement that the NHEK region is glued to a Kerr region at $r \ra \infty$ more precise (with $\lambda \ra 0$ per definition of the NHEK metric). 
We propose that the radial energy outflux in \eqref{ErKerr} evaluated at the gluing radius $r_*$, 
\ali{
	\mathcal E^{\hat r}(r_*)= -\frac{\lambda^2}{2} \left. \left( T^r_{\ph{\mu} t} - \frac{1}{\lambda} T^r_{\ph{\mu} \phi} \right)\right\rvert_{r=r_*} ,    \label{Prplots}
}
gives a good measure for determining whether a force-free electrodynamics solution in a near-horizon geometry is energy-extracting. What is relevant in particular, is the sign of the total energy outflux through a constant $\hat r$ hypersurface $\Sigma_{\hat r}$, 
\ali{
	E  = \int_{\Sigma_{\hat r}} \mathcal E^{\hat r}(r_*)  \, d\Sigma_{\hat r}    \label{ourenergy}
}
with the directed surface element $d\Sigma_{\hat r}$ in Kerr related to the one in NHEK by $d\Sigma_{\hat r} = \frac{1}{2M^2 \zeta} d\Sigma_r$. For the axisymmetric and stationary 
solutions of section \ref{sectionsols}, the flux \eqref{Prplots} is a function of $\hat \theta$ only, 
and as $d\Sigma_r$ is just the volume element $dt \wedge d\phi \wedge d\theta$ multiplied with a positive function of $\theta$, it suffices for the determination of the sign of the extracted energy during a time interval $\Delta \hat t$ to evaluate 
\ali{
	E \sim 2 \pi \Delta \hat t \int_0^{\pi/2} \mathcal E^{\hat r}(r_*) \, d\hat \theta \, . 	
}

Our proposal can be compared to the energy outflux measured by a  zero angular momentum observer (ZAMO) in the NHEK frame. For a constant $t$ surface in NHEK with unit normal $u_\mu \sim (-r,0,0,0)$, the associated ZAMO $u^\mu \sim (\frac{1}{r},-1,0,0)$ is such that upon evaluation at $r_*$ it takes the form 
\ali{ 
	u^\mu(r_*)  \sim  (\zeta, -\frac{1}{2M},0,0) \,\, .  \label{uZAMO}
}  
By comparison with the rotating NHEK observer $\chi^\mu$ in \eqref{rotNHEKobs}, we conclude that the energy outflux measured by the ZAMO $u^\mu(r_*)$ matches (up to possible normalizations) the one in \eqref{Prplots} measured by the stationary asymptotic Kerr observer $\chi^{\hat \mu} = (1,0,0,0)$. This argument mimics the comments \eqref{eqB6} and \eqref{eqB7} in Appendix \ref{appCurvedEM} as applied to the toy model. 

\subsection{Energy flux in AdS-Kerr}

The discussion in the previous subsection can straightforwardly be repeated for the AdS case. 
As summarized in Appendix \ref{appAdSNHEK}, 
the transformation between AdS-Kerr (hatted coordinates) and its near-horizon region AdS-NHEK 
is   
\ali{
	\hat t = t \frac{r_0}{\epsilon},  \qquad \hat \phi = \phi + \Omega_H^{ext} \frac{t \, r_0}{\epsilon}, \qquad \hat r = r_+ + \epsilon \, r_0 \, r ,  \qquad \hat \theta = \theta ,  \label{coordtransfAdS}
}	
with $\Omega_H^{ext}$ given in \eqref{OmegaHextAdS} and $r_0$ in \eqref{defr0}.  The same coordinate transformation applies for the SE-NHEK limit from super-entropic AdS-Kerr, if $\hat \phi$ in \eqref{coordtransfAdS} refers to the angle $\hat \psi$ in the metric \eqref{eq:SEKerr}, and with $\Omega_H^{ext}$ given in \eqref{OmegaHSE} and $r_0$ in \eqref{r0SE}. For the rest of this section we will refer to AdS-NHEK for definiteness, but the whole discussion applies equally well to SE-NHEK.  

The conserved energy flux \eqref{Ehatdef} measured by an observer $\chi^{\hat \mu} = (1/\Xi,0,0,0)$ 
at rest in AdS-Kerr is given in terms of the stress tensor components of the near-horizon observer 
\ali{
	\chi^\mu \sim \left(\frac{\epsilon}{r_0}, -\Omega_H^{ext},0,0 \right) \label{chiAdS} 
} 
as 
\ali{
	\mathcal E^{\hat r}
	&= -\epsilon r_0 \left( \frac{\epsilon}{r_0} T^r_{\ph{\mu} t} - \Omega_H^{ext} T^r_{\ph{\mu} \phi} \right) \,\, .  \label{ErKerrAdS}
}
The scaling parameter 
\ali{
	\lambda = \frac{\epsilon}{r_0 \Omega_H^{ext}} 
		\label{lambda-AdS}
}
sets the maximum value $1/\lambda$ of the rotation 
$\omega = k \, r$ of the NHEK region, resulting in a gluing radius 
\ali{
	r_* = \frac{1}{k \lambda}  \label{rstarAdS}
} 
at which we evaluate the flux,  
\ali{
	\mathcal E^{\hat r}
	(r_*) = -\epsilon r_0 \left. \left( \frac{\epsilon}{r_0} T^r_{\ph{\mu} t} - \Omega_H^{ext} T^r_{\ph{\mu} \phi} \right)\right\rvert_{r=r_*} \, . 
}
An AdS-NHEK ZAMO $u^\mu \sim (\frac{1}{r},-k,0,0)$ evaluated at $r_*$ is of the form 
\ali{ 
	u^\mu(r_*) \sim  \left(\frac{\epsilon}{r_0}, -\Omega_H^{ext},0,0 \right)  \label{AdSZAMO}
}
so that the Poynting flux measured by this observer will match the  flux $\mathcal E^{\hat r}$ measured by $\chi^\mu$.

\subsection{Discussion of energy flux profile} 

The main equations for energy extraction are given in equations \eqref{toymodel49} (for the toy model), \eqref{ErKerr} (for the Kerr energy flux $\mathcal E^{\hat r}$) and \eqref{ErKerrAdS} (for the AdS-Kerr energy flux $\mathcal E^{\hat r}$). These equations reflect the transformation under respectively the coordinate transformation between the lab and rotating frame in the toy model, and the near-horizon coordinate transformation, with $T^{\hat \mu}_{\phantom{\mu} \hat \nu} = T^\mu_{\phantom{\mu}\nu} \frac{\p x^{\hat \mu}}{\p x^\mu} \frac{\p x^\nu}{\p x^{\hat \nu}}$.  
Let us distill from the (AdS-)Kerr energy flux formula the main expected behavior. More specifically we want to determine what factors will be crucial to the sign of the flux. 

We start by repeating here the equation that can be applied for both the Kerr and the AdS-Kerr energy fluxes, i.e.~equation \eqref{Prplots} with $r_*$ defined in \eqref{rstarAdS},  
\ali{
	\mathcal E^{\hat r}(r_*) \sim  - \left. \left( T^r_{\ph{\mu} t} - \frac{1}{\lambda} T^r_{\ph{\mu} \phi} \right)\right\rvert_{r=r_*} ,   \qquad r_* = \frac{1}{k \lambda} .  \label{fluxeq}
	}
For a NHEK solution, the corresponding Kerr flux is given by this $\mathcal E^{\hat r}(r_*)$ with the scaling parameter $\lambda$ defined in \eqref{lambdaNHEK} and $k=1$.  For an AdS-NHEK solution, similarly the AdS-Kerr flux is given by the same equation with $\lambda$ now equal to \eqref{lambda-AdS}, with parameters $r_0$ and $\Omega_H^{ext}$ given in \eqref{defr0} and \eqref{OmegaHextAdS}, and $k$ defined under equation \eqref{eq:AdsDefs}. Finally, for an SE-NHEK solution, one can again use the flux equation with \eqref{lambda-AdS}, where now $r_0$ and $\Omega_H^{ext}$ are the ones from equations \eqref{r0SE} and \eqref{OmegaHSE}, and $k$ is defined in \eqref{eq:SEdefd}.  

The two terms in the above flux equation are not independent, but related to each other as 
\ali{
	T^r_{\phantom{r}\phi} = - \frac{1}{\omega_F} T^r_{\phantom{r}t}.  
}
Plugging this relation into \eqref{fluxeq} and replacing $\lambda$ by $k r_*$, we can write 
\ali{
	\mathcal E^{\hat r}(r_*) \sim  - \left. T^r_{\ph{\mu} t} \left( 1 + \frac{k r}{\omega_F} \right) \right\rvert_{r=r_*} \, . 
}
All the geometries we consider are of the form \eqref{NHE}, which in turn is of the form of a general rotating metric 
\eqref{eq:metric2} with angular velocity $\Omega \equiv -g_{t\phi}/g_{\phi\phi}$ equal to $\Omega = -k r$. This allows to further rewrite the flux profile as 
\ali{
	\mathcal E^{\hat r}(r_*) \sim  \left. -\omega_F \, \mathcal E^r_{rot} \, \left( \Omega - \omega_F  \right) \right\rvert_{r=r_*} ,  \label{fluxsign}  
}
where we used the notation $\mathcal E^r_{rot} \equiv -T^r_{\phantom{r}t}$ for the flux measured by an observer at rest in the near horizon region or `rotating frame'. 
It is clear from this form that the sign of the flux will be determined by whether or not the metric angular velocity is larger than the field angular velocity, as well as by the sign of the field angular velocity $\omega_F$ and the sign of the flux as measured by a non-rotating observer in the near-horizon region. 

Equation \eqref{fluxsign} for the (AdS-)Kerr flux makes the equivalence with the toy model manifest: coupling back to the notation used in section \ref{toymodelsection} by replacing $\mathcal E^{\hat r} \ra P^r$, $\mathcal E^r_{rot} \ra P^r_{rot}$ and $\Omega \ra \Omega_H$, equation \eqref{fluxsign} expresses that the lab Poynting flux $P^r$ scales like $-\omega_F P^r_{rot} (\Omega_H - \omega_F)$, as can be read off directly from the expressions in Figure \ref{figtoymodel}.

\subsection{Energy flux in BTZ} 
\label{subsec:BTZ} 

The aim here is to demonstrate the extent of the validity of applying the proposed toy model to the derivation of an energy flux from extreme near horizon black hole throats. 
To this end we employ an exact solution to FFE for rotating black holes, namely, the solution describing the BZ process \cite{Jacobson:2017xam} in the 3-dimensional rotating black hole  known as BTZ. The presumably unique, exact EM field is given in a  simple, closed form for the BTZ metric \eqref{ds2BTZ} with coordinates $(\hat t, \hat \phi,\hat r)$. This result provides us with a set-up where the \emph{full} FFE solution is known, i.e.~not just in the near-horizon region. In particular, the FFE solution in BTZ will allow us to check equation \eqref{fluxBTZ} for the BTZ energy flux as measured by a NHEBTZ observer. 

We begin by directly applying the near horizon procedure to the FFE solution in the BTZ metric. 
The energy extracting FFE solution in BTZ obtained in  \cite{Jacobson:2017xam} is given by 
\ali{
	\hat F = \frac{\dot \Phi}{2 \pi } d\hat t \wedge d\hat \phi + \frac{Q}{2 \pi \hat  r} d\hat r \wedge d\hat t - \frac{Q \, \hat r (\Omega (\hat r)-\Omega_F)}{2 \pi  \alpha(\hat r)^2} (d\phi - \Omega(\hat r)  d\hat t) \wedge d\hat r  \label{hatFBTZ}
}
with electric charge $Q$ and magnetic monopole current $\dot \Phi$, and the functions $\alpha$ and $\Omega$ given in \eqref{alphaOmegaBTZ}. Including Znajek's regularity condition implies $\dot \Phi= Q r_+(\Omega_H-\Omega_F)$, where the horizon angular velocity is given by $\Omega_H=r_-/(l r_+)$ in terms of the horizons $r_\pm$. 

Consider in particular the extreme BTZ solution ($\Omega^{ext}_H=1/l$), with $\dot \Phi = 0$ (required to have a finite field strength in the NHEBTZ background) and $\Omega_F = 1/l$. Its energy flux as defined in \eqref{Ehatdef}, with $\chi = \p_{\hat t}$, reads 
\ali{
	\mathcal{E}^{\hat\mu} = \frac{Q^2}{4\pi^2 (\hat r^2 - r_+^2)} \left(1,\frac{1}{l},0\right)  \, .  \label{EmuhatBTZ}
} 
Note that there is no radial energy outflux for this special case of the extreme BTZ solution. This is a consequence of the regularity condition imposed by the near-horizon solution. It causes the field angular velocity to reach the horizon angular velocity of the extremal BTZ black hole, $\Omega_F = \Omega_H^{ext}$, 
while energy extraction is achieved in the range $0 < \Omega_F < \Omega_H^{ext}$. 

Following the near horizon limiting procedure defined in Appendix \ref{appBTZ} (which involves defining new coordinates $x^\mu$ in addition to taking the limit $\epsilon \rightarrow 0$) for the energy flux $\mathcal E \equiv \mathcal E_{\hat{\mu}} dx^{\hat{\mu}}= \mathcal E_\mu dx^\mu=\frac{Q^2}{4 \pi^2 l} d\phi$ with (\ref{EmuhatBTZ}) yields
\ali{\label{eq:mathcalEBTZ}
\mathcal E^\mu = \left( \frac{Q^2}{4 \pi^2 r},0,0 \right),
} 
and for the extreme electromagnetic field (\ref{hatFBTZ}) one finds
\ali{ \label{NHEBTZField}
	F = \frac{l Q}{4 \pi r} d\phi \wedge dr \, . 
}
This result for the near horizon solution in the NHEBTZ background \eqref{ds2nhBTZ} 
is obtained specifically by making use of the conditions $\dot \Phi = 0$ and $\Omega_F = 1/l$ which impose finiteness and Znajek's condition. 

Imagine we had not known the full FFE solution in BTZ, but only the field \eqref{NHEBTZField} in NHEBTZ. The only non-vanishing component of the corresponding stress energy tensor is $T^{t}_{\phantom{t}\phi}=\frac{l Q^2}{4 \pi^2 r}$. Following the discussion in the preceding sections, the energy flux measured by a rotating NHEBTZ observer $\chi = \epsilon \p_t - \frac{1}{l} \p_\phi$ yields
\ali{
	\mathcal E^\mu = -\left( \epsilon {T^{\mu}}_{t} - \frac{1}{l} {T^{\mu}}_{\phi} \right) = \left( \frac{Q^2}{4 \pi^2 r},0,0 \right),  \label{EmuNHEBTZ}
}
in agreement with (\ref{eq:mathcalEBTZ}). The energy flux (\ref{EmuNHEBTZ}) computed solely with the field in NHEBTZ gives rise to the same energy flux $\mathcal{E}$ as the one measured by the $\chi = \p_{\hat t}$ observer at rest in BTZ. This is the extent to which our analysis of the NHEBTZ is exactly equivalent to our toy model prescription. 

Now let us take a look at the gluing procedure. It suggests that the object 
\ali{
	\mathcal E^{\hat\mu}(r_*) = - \left. \frac{\p x^{\hat \mu}}{\p x^\mu}\left( \epsilon {T^{\mu}}_{t} - \frac{1}{l} {T^{\mu}}_{\phi} \right) \right|_{r = r_*} = \frac{Q^2}{4 \pi^2 r_* \epsilon} \left( 1,\frac{1}{l},0 \right)  \label{fluxBTZ}
} 
allows the NHEBTZ observer, with no knowledge of the full BTZ solution, to obtain the profile of the energy outflux measured outside the throat. 
The gluing radius $r_*$ marking the `edge' of the near horizon region is determined to be 
\ali{
	r_* = \frac{r_+^2}{\epsilon}.  
	} 
This follows from a completely analogous reasoning as before. Namely, the angular velocity $\Omega_A$, which can be read off from the BTZ to near-horizon angle transformation $\hat \phi = \phi + \Omega_A t$ given in \eqref{nhBTZcoord}, sets the maximal value $\omega(r_*)$ of the rotation 
$\omega \equiv g_{t\phi}/g_{\phi\phi} = \frac{r}{l r_+^2}$ of the NHEBTZ metric \eqref{ds2nhBTZ}.  
Writing out the evaluation at $r_*$ in \eqref{fluxBTZ} gives rise to a finite quantity 
\ali{ 
	\mathcal E^{\hat \mu}(r_*) = \frac{Q^2}{4 \pi^2 r_+^2 } \left( 1, \frac{1}{l},0 \right). 
} 
It matches the full BTZ flux in \eqref{EmuhatBTZ} evaluated at the gluing point $\hat r_* = \sqrt{r_+^2 + \epsilon r_*}$.


\section{Energy flux of force-free solutions}  \label{sectionwithfigs}

We are interested in the profile of $\mathcal E^{\hat r}(r_*)$ (defined in the previous section) as a function of $\hat \theta$ ranging from the north pole to the equator, and the positivity of $\int_0^{\pi/2} \mathcal E^{\hat r}(r_*) d\hat \theta$ as a sign of energy extraction. 
The Kerr flux  $\mathcal E^{\hat r}$ takes the role of the Poynting vector in the lab frame of the toy model of section \ref{toymodelsection}. It can be energy extracting even if the NHEK  flux, analogous to the Poynting vector in the rotating frame of the toy model, is not. To analyze this statement, we will also consider the NHEK flux 
as measured by a stationary asymptotic observer $(1,0,0,0)$ in the NHEK frame, $- T^r_{\phantom{r}t} $, and the corresponding extracted energy at the boundary of NHEK 
\ali{
	E_{NH} = -\lim_{r \ra \infty} \int_{\Sigma_{r}}  T^r_{\phantom{r}t} \, d\Sigma_{r}  , 
} 
with $d\Sigma_{r} = \sqrt{-g} \, dt \wedge d\phi \wedge d\theta$. 
This is the object considered in \cite{Lupsasca:2014pfa} to discuss energy extraction\footnote{
	To compare to their notation, $\lim_{r \ra \infty} T^r_{\phantom{r}t} \,d\Sigma_{r}$ becomes their $\mathcal E_\infty = \sqrt{-h} \,  n^\nu_{(r)} T_{\alpha\nu} \, \chi^\alpha_{(t)}|_{r \ra \infty}$ (with $\chi = \p_t$ and $n$ the normal to $\Sigma_r$). It is pointed out in their footnote 2 that the NHEK and Kerr energies differ because of the mixing with angular momentum.   
}. 
It is to be compared to the energy \eqref{ourenergy}, rewritten using equation \eqref{ErNHEK} 
as 
\ali{
	E  = - \lim_{\zeta \ra 0} \int_{\Sigma_{r}} \left( \zeta T^r_{\ph{\mu} t}(r_*) - \frac{1}{2M} T^r_{\ph{\mu} \phi}(r_*) \right) \, d\Sigma_{r} \, ,   \label{energy}
}
where now we explicitly wrote the limit $\zeta \ra 0$ that was implicit in equation \eqref{ourenergy}, and $r_* = 1/(2 M \zeta)$. The AdS version of this expression is immediate from \eqref{chiAdS}: 
\ali{
	E  = - \lim_{\epsilon \ra 0} \int_{\Sigma_{r}} \left( \frac{\epsilon}{r_0} T^r_{\ph{\mu} t}(r_*) - \Omega_H^{ext}  T^r_{\ph{\mu} \phi}(r_*) \right) \, d\Sigma_{r} \, ,  
}  
with $r_* = r_0 \Omega_H^{ext} / (k \epsilon)$. 

For the scaling solutions (both in NHEK and AdS-NHEK), the scaling of the relevant stress tensor components is  $T^r_{\phantom{r}t} \sim r^{2h+2}$ and $T^r_{\phantom{r}\phi} \sim r^{2h+1}$. It follows that the extracted energy \eqref{energy} scales like  
\ali{
	E \sim \lim_{\zeta \ra 0} \, 2 \pi \Delta t \,\, r_*^{2 h + 1}  
}
where we used the NHEK notation, but the same conclusion will apply to the AdS-NHEK case. 
The NHEK energy flux rate $E/\Delta t$ for the scaling solution will thus be finite only for the special value $h=-\frac{1}{2}$. This special value was also pointed out in \cite{Lupsasca:2014pfa}\footnote{
	Our $h$ differs from the one in \cite{Lupsasca:2014pfa} by a sign. 
}. However, from the point of view of the Kerr frame, the energy flux is measured in units of Kerr time $\hat t = t/\zeta$, 
\ali{
	E \sim \lim_{\zeta \ra 0} \,  2 \pi \Delta \hat t \,\, r_*^{2 h}  
}
and the Kerr energy flux rate $E/\Delta \hat t$ is finite for $h=0$. This motivates the study of the log ansatz discussed in section \ref{sectionlogansatz}. Indeed the log ansatz allows for a finite energy outflux in this sense: it has 
$T^r_{\phantom{r}t} \sim r^2$ and $T^r_{\phantom{r}\phi} \sim r$ or $\mathcal E^{\hat r}$ finite. 

We discuss now in detail the $h=1$ scaling solutions in NHEK and AdS-NHEK. They have infinite energy outflux. 

The $h=1$ scaling solution in NHEK was previously discussed in \cite{Zhang:2014pla} (and \cite{Lupsasca:2014hua}). We repeat it here in our notation before we present the new solutions: the AdS-NHEK $h=1$ scaling solution and an SE-NHEK log solution.

\paragraph{Scaling solution in NHEK} 

\begin{figure}[t] 
\centering  \includegraphics[width=8cm]{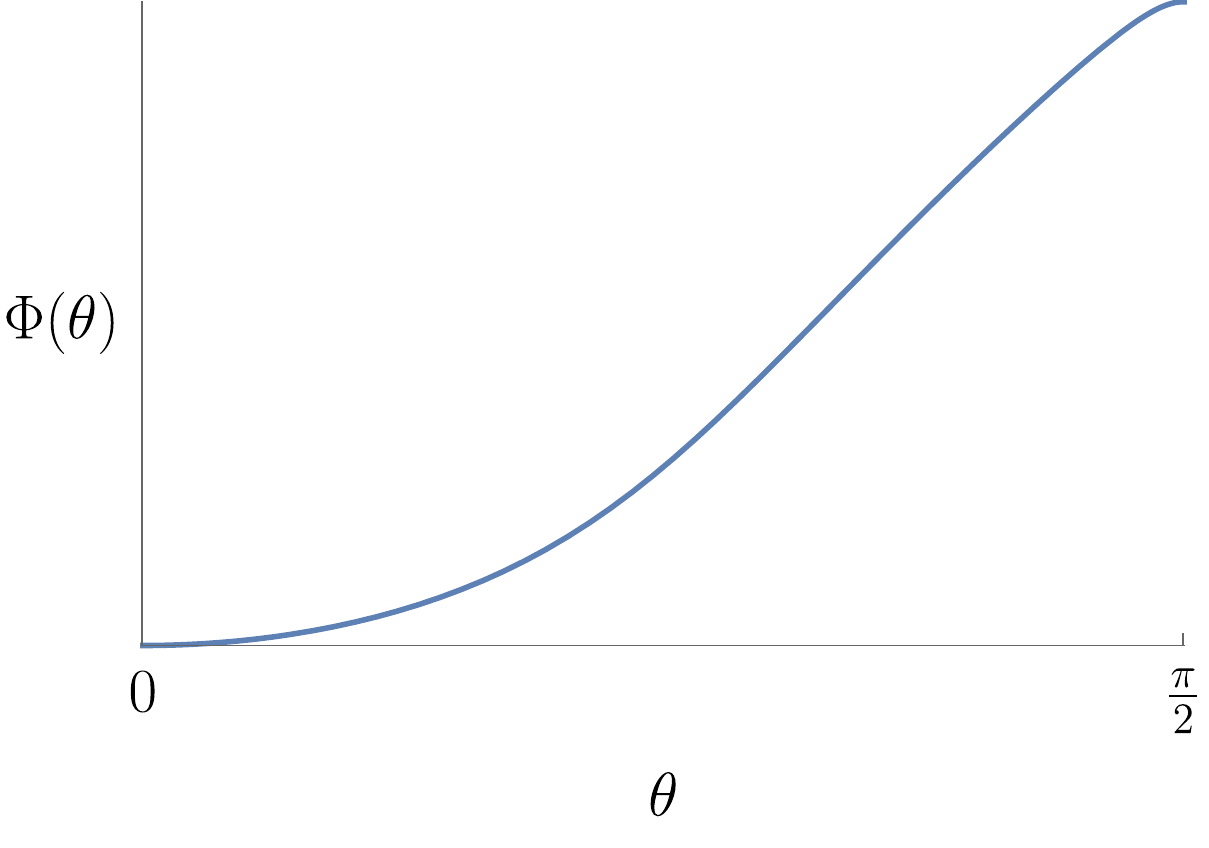} \\ 
	\centering \includegraphics[]{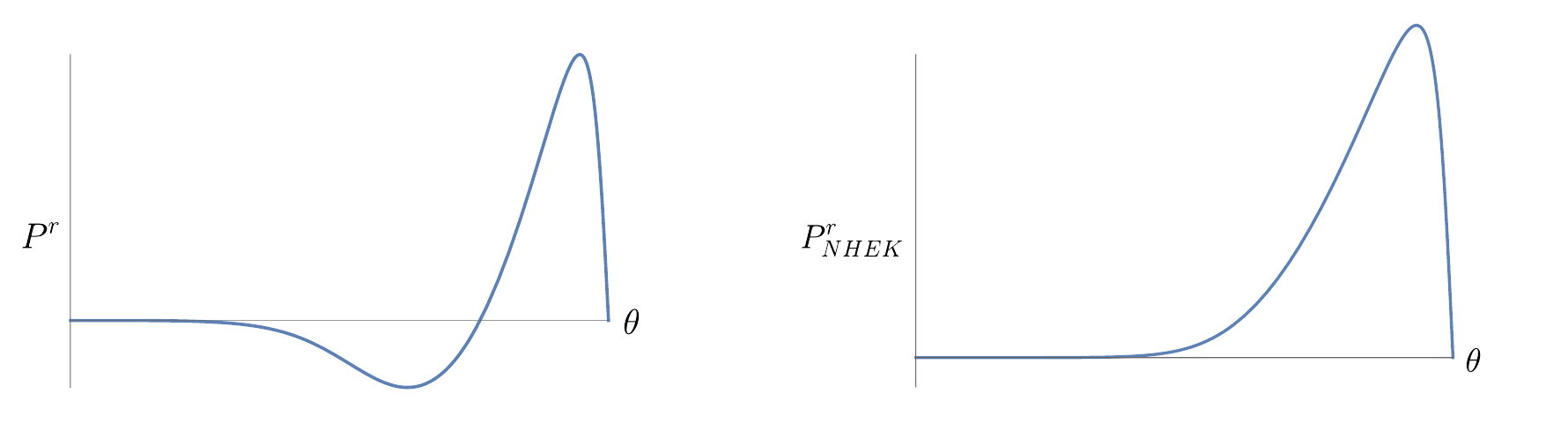} 
	\caption{Solution $\Phi(\theta)$ and corresponding Poynting flux $P^r$ as a function of $\theta$, for the energy extracting NHEK scaling solution for parameter values 
	$\omega_0 = -1, h = 1, I_0 = 0.1$ and $\Omega_H^{ext} = \frac{1}{2M}=0.5$, and with $\Phi(\pi/2) =1.44774$. This reproduces the solution of \cite{Zhang:2014pla} (up to a sign disagreement). The Kerr flux is smaller than the NHEK flux.} \label{figscalingNHEK}
\end{figure}

For the scaling ansatz in section \ref{sectionscaling}, we find  NHEK and Kerr stress tensor components 
\ali{
	T^r_{\phantom{r}t} = \frac{\omega_0 I_0 r^{2h+2}  \Phi(\theta)^{1 + 2/h}\Phi'(\theta)}{4 M^4 \Gamma(\theta)^2 \Lambda(\theta)}, \qquad -T^r_{\phantom{r}\phi} =\frac{T^r_{\phantom{r}t}}{\omega_F} = \frac{ T^r_{\phantom{r}t} }{r \omega_0 \Phi^{1/h}}\label{Tscaling}
}
and 
\ali{
	\mathcal E^{\hat r} =  -2 M^2 \zeta \left( \zeta T^r_{\ph{\mu} t} - \frac{1}{2M} T^r_{\ph{\mu} \phi} \right) = \frac{I_0 r^{2h+1} \zeta \Phi(\theta)^{1+1/h} \left( 1 + 2 \omega_0 M r \zeta \Phi(\theta)^{1/h} \right) \Phi'(\theta)}{4 M^3 \Gamma(\theta)^2 \Lambda(\theta)} \, . \label{Thatscaling}
}
The first plot in Figure \ref{figscalingNHEK} shows the numerically obtained solution $\Phi(\theta)$ of the EOM \eqref{NHEKscalingEOM} for $h=1$. It satisfies the boundary conditions \eqref{bc1} and \eqref{bc2}. What is plotted in the second figure is the infinite energy flux rate $\mathcal E^{\hat r}(r_*)$ rescaled by a power of $\zeta$ to extract the $\theta$-profile of the Poynting flux, 
\ali{
	P^r = \zeta^{2h} \mathcal E^{\hat r}(r_*), \qquad r_* = \frac{1}{2 M \zeta} \, .     
}
This Kerr outflux is compared in the third plot to the (infinite) NHEK outflux rescaled to the finite 
\ali{
	P^r_{NHEK} = -\zeta^{2h+2} T^r_{\phantom{r}t}(r_*), \qquad r_* = \frac{1}{2 M \zeta} \, .    
}
Both profiles have a positive integral over $\theta$, i.e.~both $E$ and $E_{NH}$ are positive, and the extracted energy $E_{NH}$ is actually greater than $E$ in this case.

This solution was first presented in \cite{Zhang:2014pla}. The Poynting flux considered in \cite{Zhang:2014pla} is the one observed by a ZAMO $u_\mu = -\sqrt{2 \Gamma(\theta)} M (r,0,0,0)$ in the NHEK frame, using the general definitions \eqref{Ecurvedspacedef}-\eqref{Pcurvedspacedef}. The radial  component for the Poynting flux \eqref{Pofudef} equals $-T^r_{\phantom{r}\rho} u^\rho$. It was discussed in the paragraph around equation \eqref{uZAMO} to produce the same energy outflux profile as we find for $\mathcal E^{\hat r}(r_*)$.

\paragraph{Scaling solution in AdS-NHEK}

\begin{figure}[t] 
	\centering \includegraphics[width=8cm]{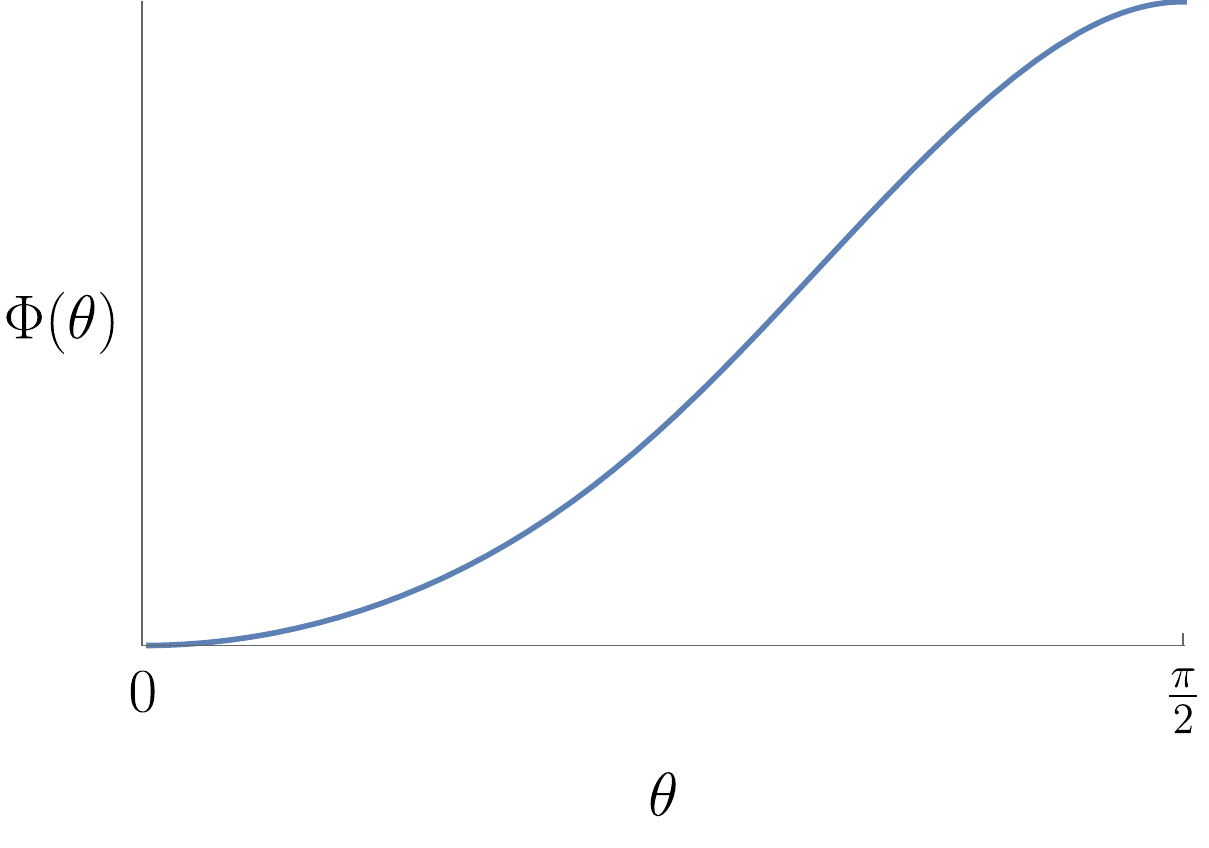} \\ 
	\centering \includegraphics[]{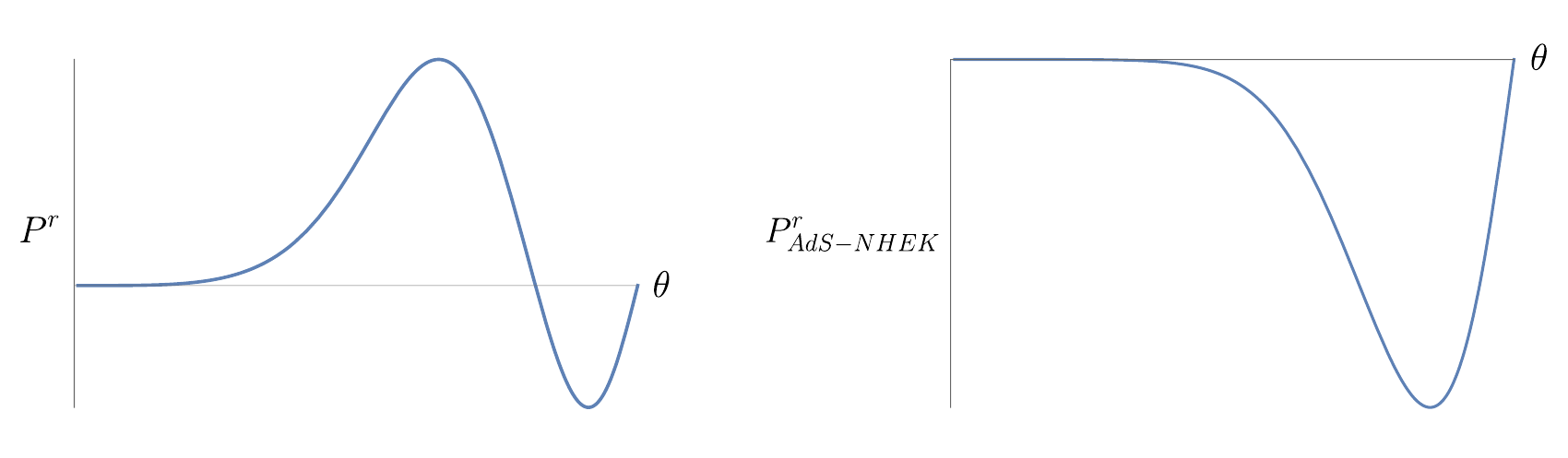} 
	\caption{Solution $\Phi(\theta)$ and corresponding Poynting flux $P^r$ as a function of $\theta$, for the AdS-NHEK scaling solution for parameter values $h = 1, I_0 = 0.2$, $\ell=1, a=-0.25, k=-0.68, \omega_0 = \frac{1}{2}, r_+ = 0.226654$ and   $\Omega_H^{ext} = -2.06$, and with $\Phi(\pi/2) = 1.585633$. The near horizon flux $P^r_{AdS-NHEK}$ is inwards while $P^r$ does detect energy extraction.  
	} 
\end{figure} \label{figAdSNHEKscaling}

For the AdS-NHEK scaling solution, we find the following AdS-NHEK and AdS-Kerr mixed stress tensor components 
\ali{
	T^r_{\phantom{r}t} = \frac{\omega_0 I_0 r^{2h+2} \Phi(\theta)^{1 + 2/h}\Phi'(\theta)}{\alpha(\theta) \gamma(\theta) \Gamma(\theta)^2}, \qquad -T^r_{\phantom{r}\phi} =
	\frac{T^r_{\phantom{r}t}}{\omega_F} =
	 \frac{T^r_{\phantom{r}t}}{r \omega_0 \Phi^{1/h}}  \label{TscalingAdS}
}
and 
\ali{
	\mathcal E^{\hat r} =  -\epsilon r_0 \left( \frac{\epsilon}{r_0} T^r_{\ph{\mu} t} - \Omega_H^{ext} T^r_{\ph{\mu} \phi} \right) = \frac{I_0 r^{2h+1} \epsilon \Phi(\theta)^{1+1/h} \left(r_0 \Omega_H^{ext} + \omega_0 \, r  \epsilon \Phi(\theta)^{1/h} \right) \Phi'(\theta)}{\alpha(\theta)  \gamma(\theta) \Gamma(\theta)^2 } \, . \label{ThatscalingAdS}
}
Figure \ref{figAdSNHEKscaling} shows the numerical function $\Phi(\theta)$ obtained for $h=1$, which solves EOM \eqref{AdSNHEKscalingEOM} with boundary conditions  \eqref{bc1}-\eqref{bc2}. Further, the $\theta$-profile of the Poynting flux, again appropriately rescaled to 
\ali{ 
	P^r = \epsilon^{2h} \mathcal E^{\hat r}(r_*) , \qquad r_* = \frac{r_0 \Omega_H^{ext}}{k \epsilon} 
}
is compared to the rescaled AdS-NHEK outflux  
\ali{
	P^r_{AdS-NHEK} = -\epsilon^{2h+2} T^r_{\phantom{r}t}(r_*) , \qquad r_* = \frac{r_0 \Omega_H^{ext}}{k \epsilon}  \, .   
}

In this case we observe an inwards AdS-NHEK flux ($E_{NH} < 0$) but an outwards flux in AdS-Kerr ($E > 0$). This is reminiscent of the toy model, and provides an example where it matters to describe the energy extraction from the point of view of a Kerr rather than a near-horizon observer.

\paragraph{Log solution in SE-NHEK} 

While we have not succeeded in finding a log solution in (AdS-)NHEK with the boundary condition $\Phi \ra 0$ at $\theta \ra 0$, we do present such a solution in SE-NHEK. 

\begin{figure}[t] 
	\centering \includegraphics[width=8cm]{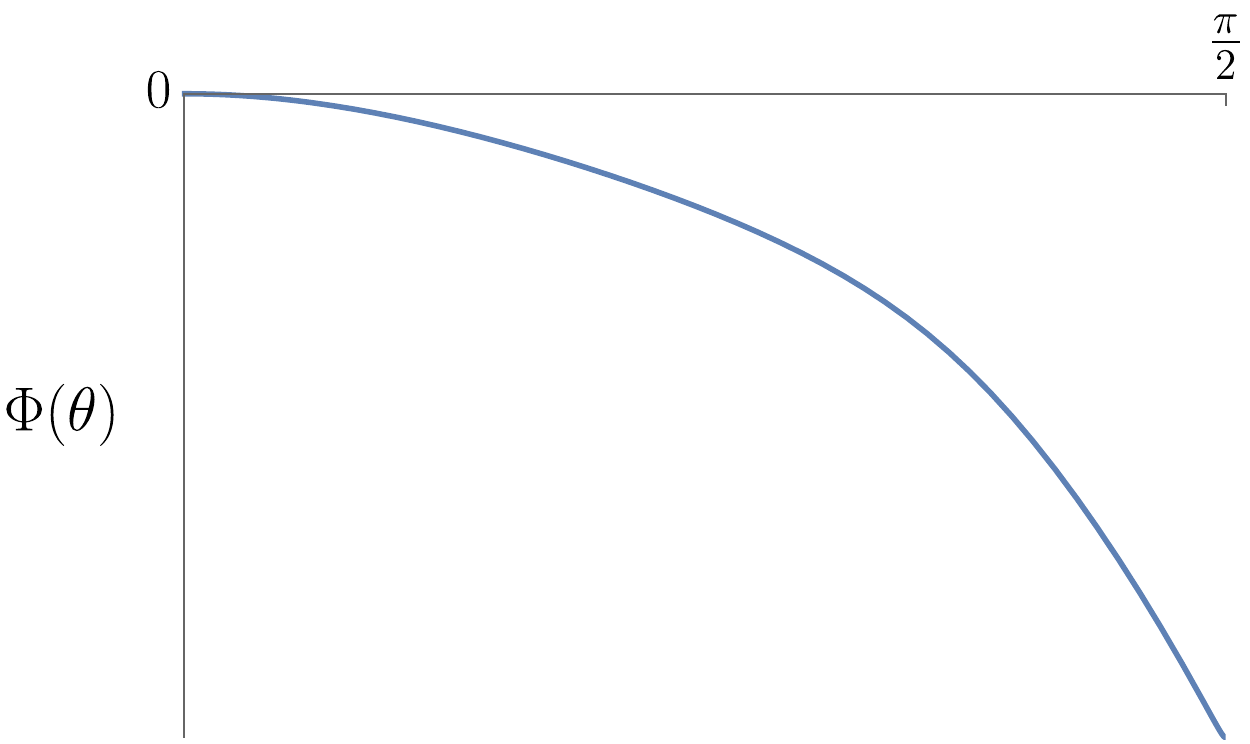}  \\ \vspace{.5cm} 
	\centering \includegraphics[]{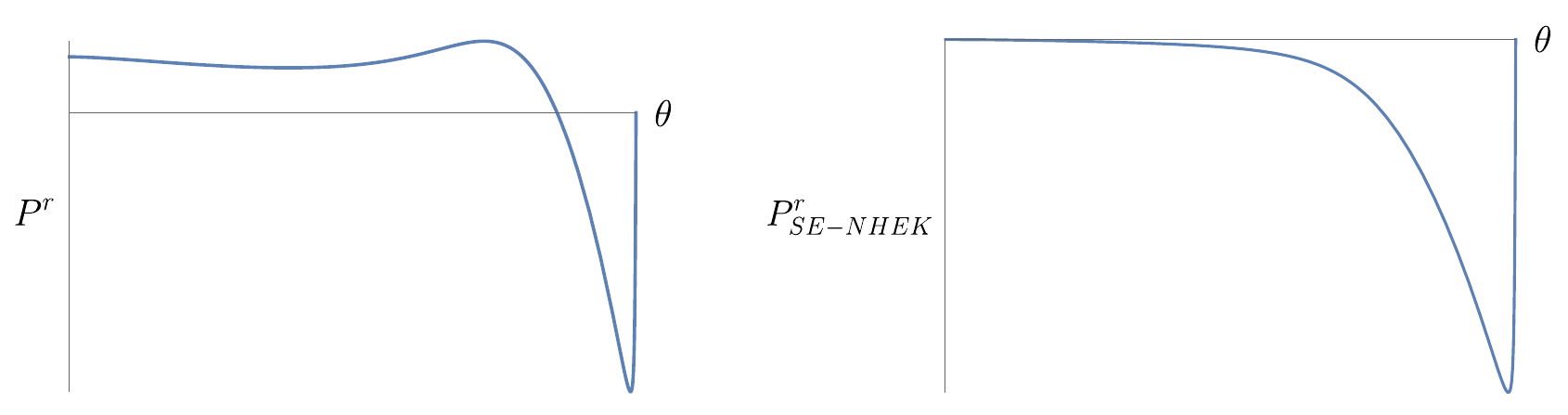}
	\caption{Solution $\Phi(\theta)$ and corresponding Poynting flux $P^r$ as a function of $\theta$, for the SE-NHEK log solution with parameter values $\ell = 1, I_0 = 0.9, \omega_0 = 
		0.5$ and $\Omega_H^{ext} = 3/4$, and with $\Phi(\pi/2) = -0.68$. The last figure shows the flux measured by a non-rotating SE-NHEK observer.}  \label{figSENHEKlogII}
\end{figure}

Figure \ref{figSENHEKlogII} shows the numerically obtained $\Phi(\theta)$ for the specified set of parameter values. 
The plotted $\Phi(\theta)$ solves the EOM \eqref{SENHEKlogEOM} with boundary conditions  \eqref{bc1}-\eqref{bc2}. 
The mixed stress tensor components 
\ali{
	T^r_{\phantom{r}t} = \frac{I_0 \, r^{2} \Phi(\theta) \Phi'(\theta)}{\alpha(\theta) \gamma(\theta) \Gamma(\theta)^2}, \qquad 
	-T^r_{\phantom{r}\phi} = \frac{T^r_{\phantom{r}t}}{\omega_F} = \frac{T^r_{\phantom{r}t}}{r \omega_0 \Phi(\theta)}  \,,
	 \label{TlogSE}
}
and 
\ali{
	\mathcal E^{\hat r} =  -\epsilon r_0 \left( \frac{\epsilon}{r_0} T^r_{\ph{\mu} t} - \Omega_H^{ext} T^r_{\ph{\mu} \phi} \right) = \frac{I_0 r \epsilon \left(r_0 \Omega_H^{ext} + r \epsilon \, \omega_0 \Phi(\theta) \right) \Phi'(\theta)}{\omega_0 \alpha(\theta) \gamma(\theta) \Gamma(\theta)^2 } \label{ThatlogSE}
}
are such that the AdS-Kerr flux rate $\mathcal E^{\hat r}(r_*)$ is finite. It is shown in Figure \ref{figSENHEKlogII} with the notation 
\ali{ 
	P^r = \mathcal E^{\hat r}(r_*) , \qquad r_* = \frac{r_0 \Omega_H^{ext}}{k \epsilon} \, , 
}
along with the (rescaled) SE-NHEK outflux  
\ali{
	P^r_{SE-NHEK} = - \epsilon^2 T^r_{\phantom{r}t}(r_*) , \qquad r_* = \frac{r_0 \Omega_H^{ext}}{k \epsilon}  \, . 
}
The SE-NHEK flux is again inwards while the AdS-Kerr flux is outwards. 

In the scaling solutions, the boundary condition $\Phi \ra 0$ at $\theta \ra 0$ gives rise to a fine-tuning of the field strength to a finite value at the north pole. For the log solution this is not the case: even though $\Phi \ra 0$, the field strength squared $F_\mn F^\mn$ diverges at $\theta \ra 0$ (so does the angular component of the Poynting vector, while $P^r$ is finite). This unphysical feature of the solution might be allowed because of the special geometry of SE-NHEK near the north pole, but this remains unclear.    

\subsection*{Discussion of energy fluxes}

The energy outfluxes $P^r$ in Figures  \ref{figscalingNHEK}-\ref{figSENHEKlogII} all have the property that they change sign at the special value of $\theta$  where the field angular velocity $\omega_F$ equals the angular velocity $\Omega$ of the geometry. This is consistent with equation \eqref{fluxsign}. The corresponding critical value  $\theta_c$ is determined from the condition $\omega_0 \Phi(\theta_c) = -1$ for NHEK, and $\omega_0 \Phi(\theta_c) =-k$ for AdS/SE-NHEK. The toy model, with constant $\omega_F$ and $\Omega_H$, has a flux of definite sign and thus no analogue of $\theta_c$.  \\

\begin{figure}[t]
	\includegraphics[angle=90,width=5.6cm]{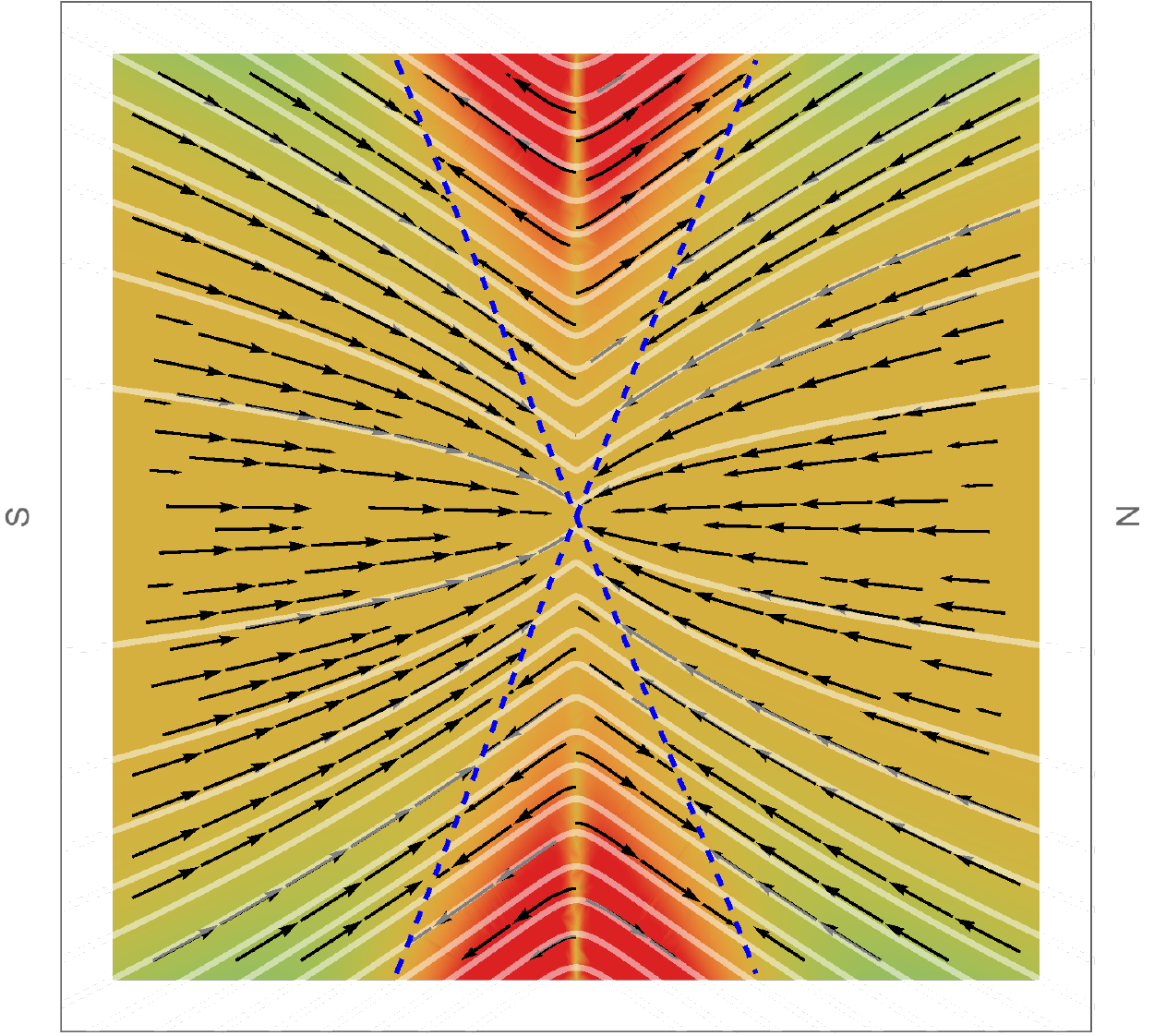}
	\includegraphics[angle=90,width=5.6cm]{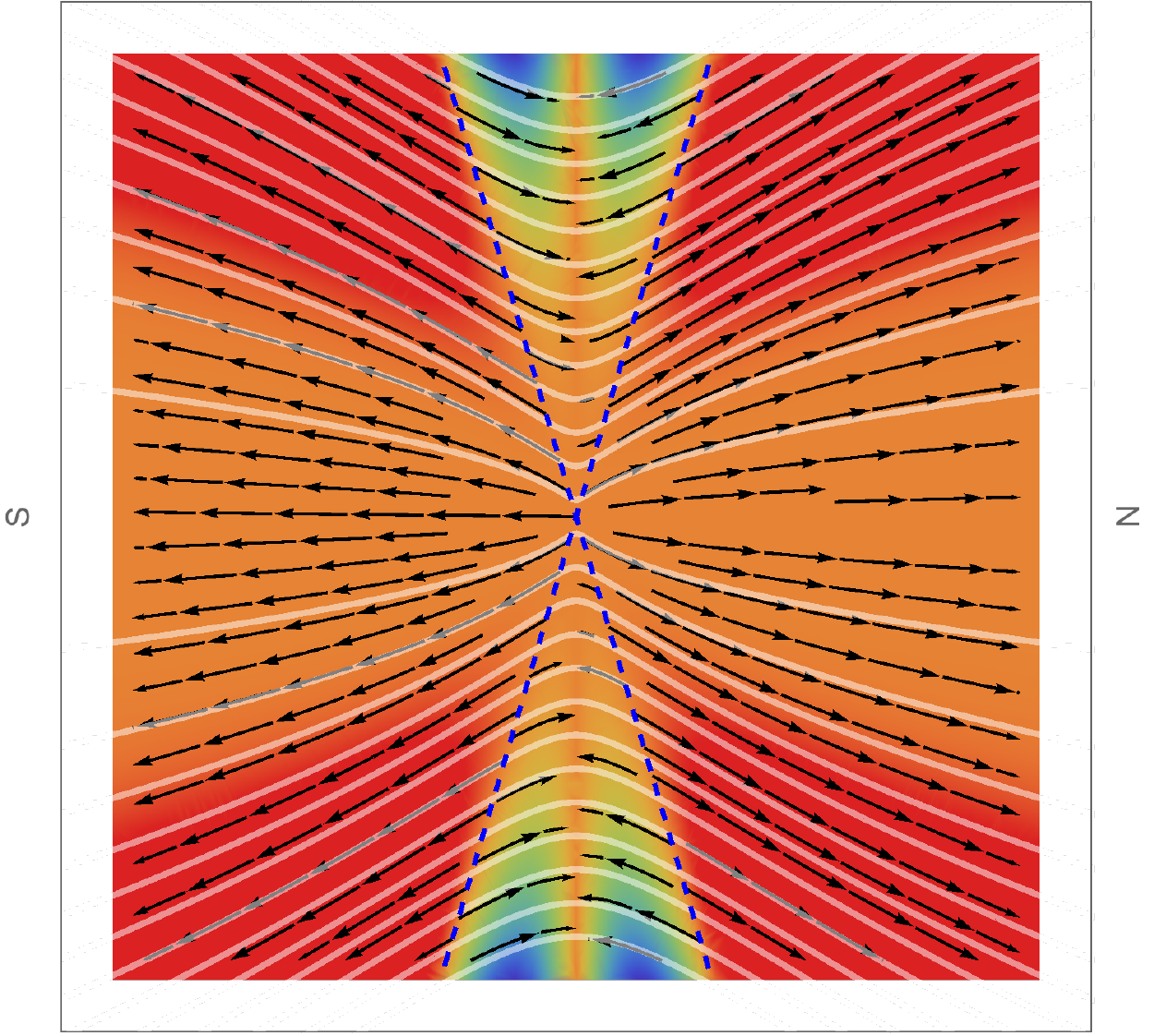}
	\includegraphics[angle=90,width=5.6cm]{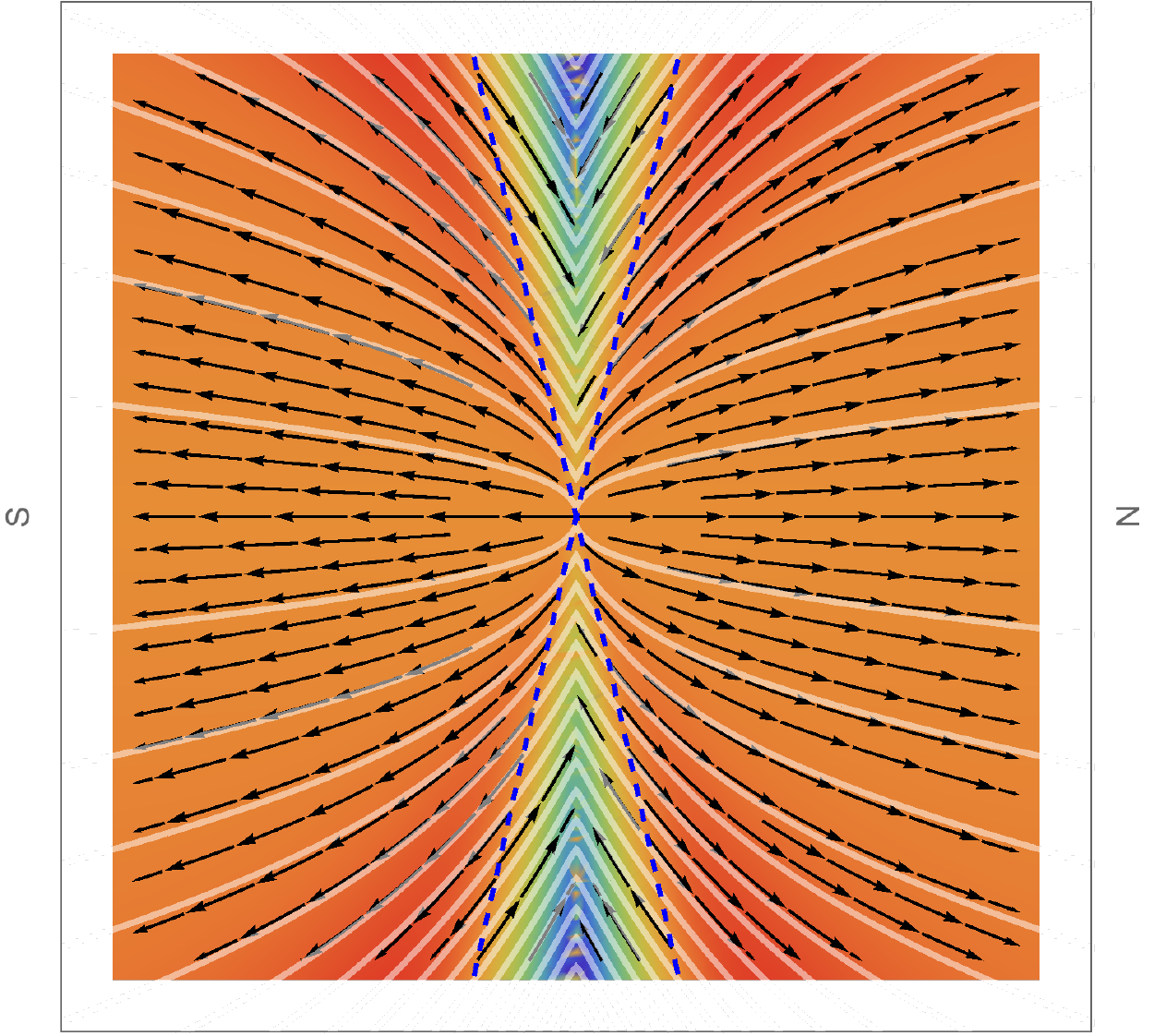}
	\caption{Polar vector plots of the (ZAMO) Poynting flux stream lines for the NHEK, AdS-NHEK and SE-NHEK solution, in the $(r,\theta)$ plane of the near horizon geometry. The flux changes sign at the critical angle $\theta_c$ (blue dashed lines). The density plot shows the magnitude of the radial Poynting flux ranging from negative (purple) over zero (yellow) to positive (red). Along the flux lines, the angular velocity $\omega_F$ is constant (white solid lines) in agreement with the conducting cylinder toy model.}
	\label{Figvectorplots}
\end{figure}

\begin{figure}[t]
	\includegraphics[width=5.6cm]{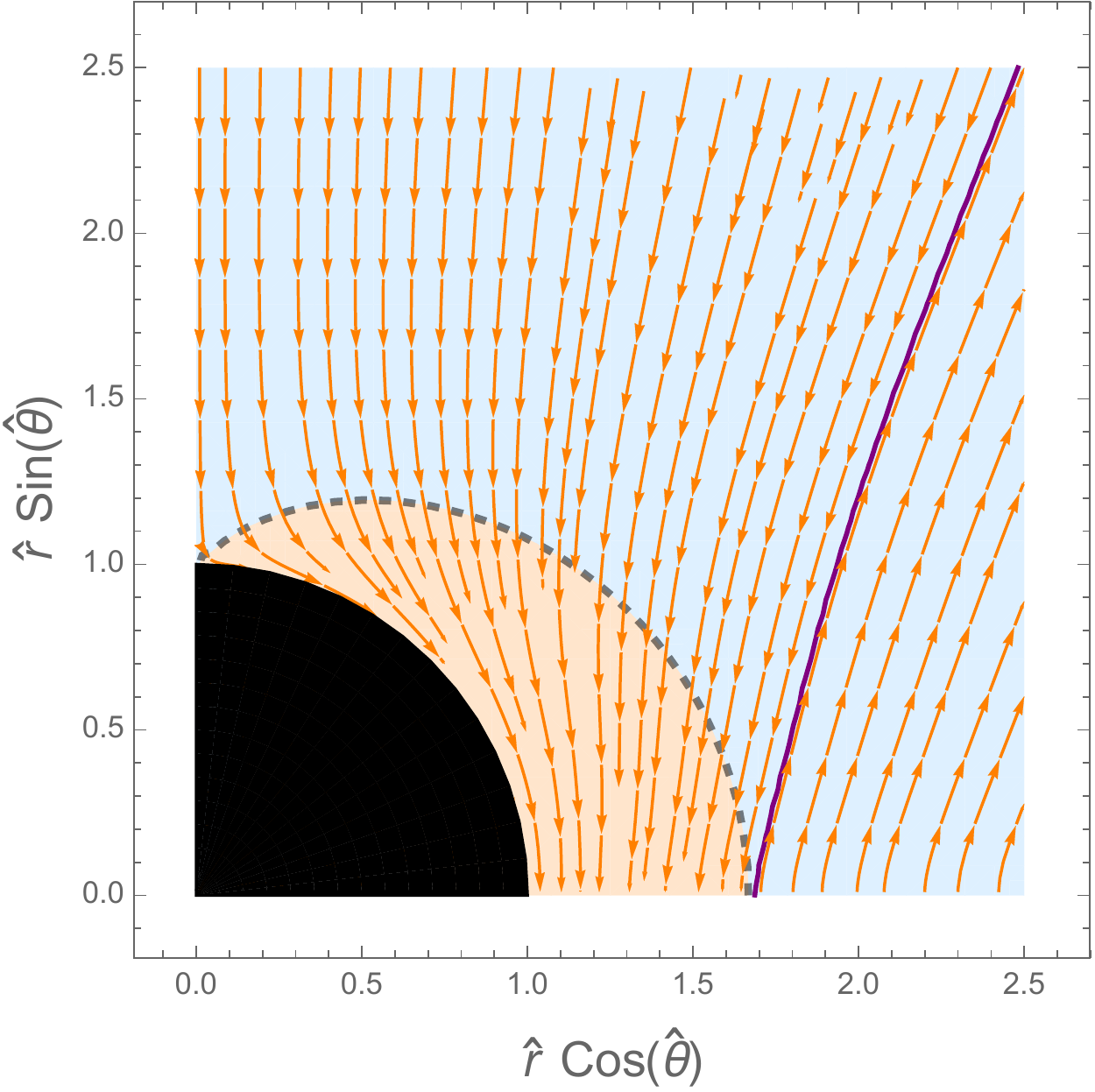}
	\includegraphics[width=5.6cm]{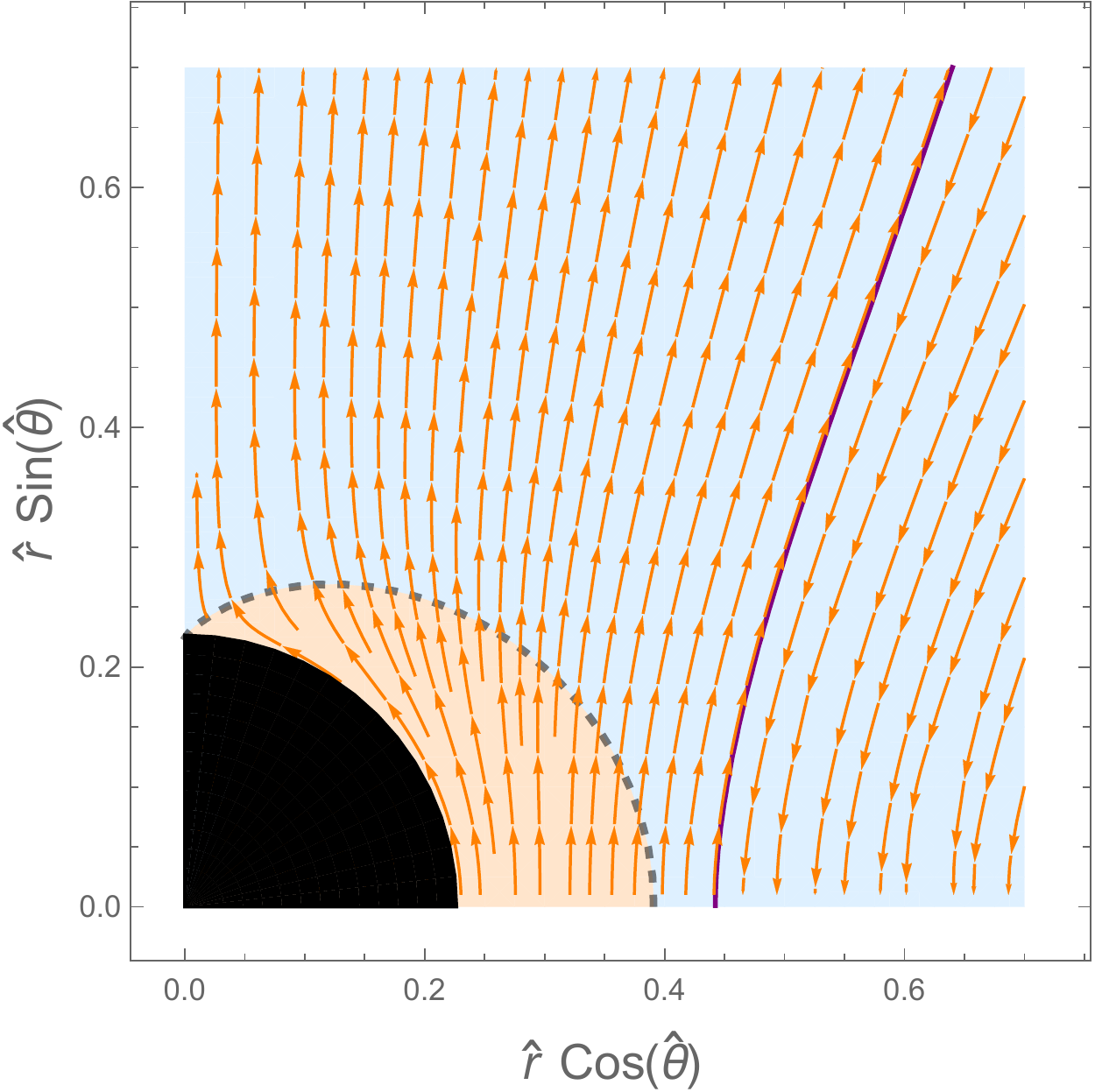}
	\includegraphics[width=5.6cm]{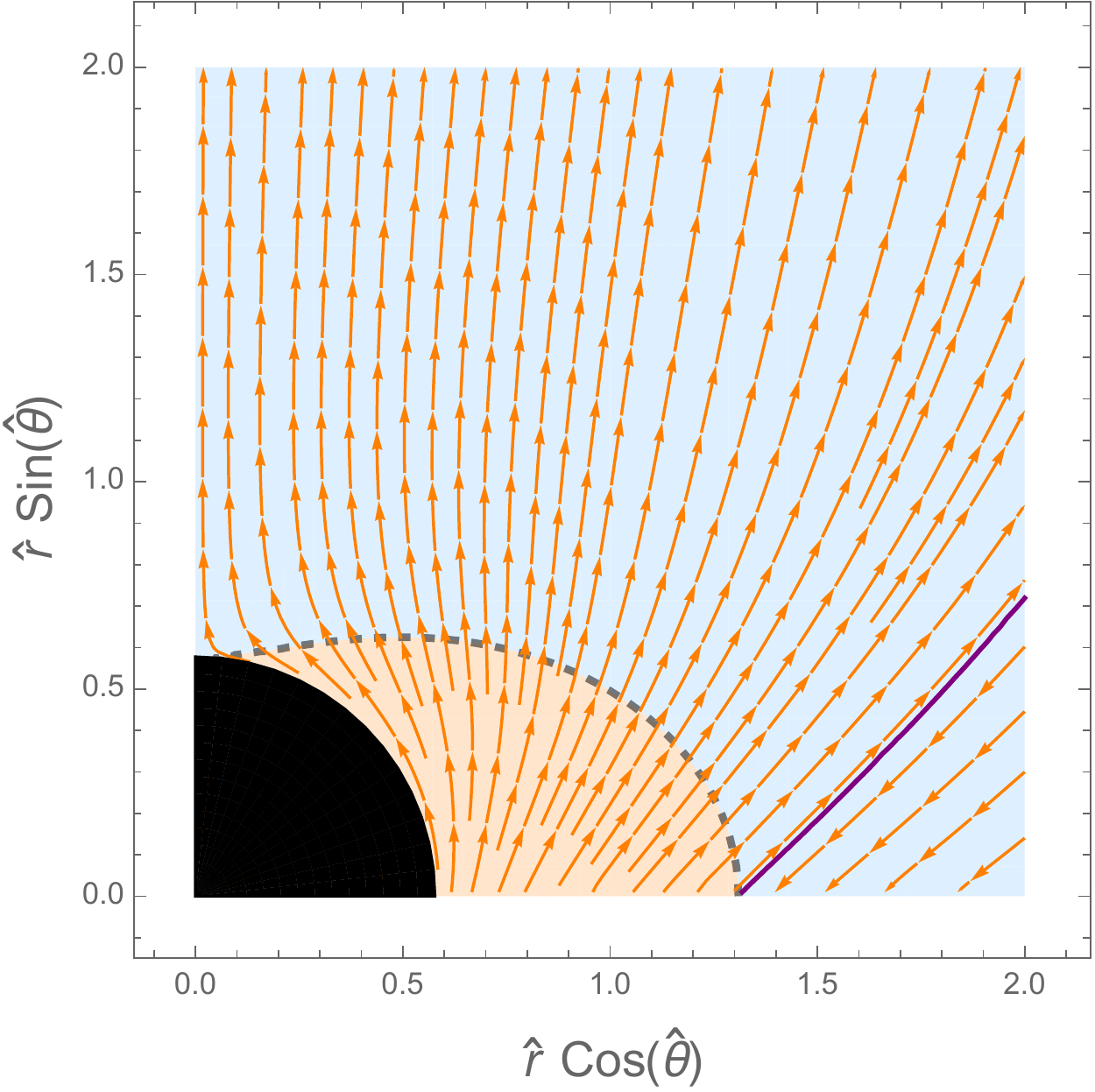} 
	\caption{Polar vector plots of the Poynting flux stream lines for the NHEK, AdS-NHEK and SE-NHEK solution, in the $(\hat r, \hat \theta)$ plane of (AdS/SE-)Kerr. Only the first quadrant 
	is shown. In NHEK and AdS-NHEK, the location where the flux changes sign (purple line) corresponds at the equator 
	to the start of the Kerr `ergoregion' as determined by the NHEK observer, which is the orange shaded region bounded by the gray dashed line. The region behind the horizon is shaded in black.
}
	\label{FigKerrvectorplots}
\end{figure}

Let us provide a different visualization of the solutions through vector plots of the flux in the $(r,\theta)$ plane of the geometry. Figure \ref{Figvectorplots} shows, for each of the solutions, the streamlines of the vector field $(P^x,P^y)$, with  $(P^x,P^y) = \left( P^{r} \cos \theta - r P^{\theta} \sin \theta, r P^{\theta} \cos \theta + P^{r} \sin \theta\right)$ in a polar vector plot, with $x = r \cos \theta$, $y = r \sin \theta$. The Poynting flux $P^\mu$ that is used to obtain these figures is the conserved energy flux measured by the rotating observer (ZAMO) in \eqref{uZAMO} and \eqref{AdSZAMO}. The result for the vector plots is qualitatively similar to the result one would obtain from using the conserved energy flux $\mathcal E^{\hat \mu}(r,\theta;\lambda)$ 
with $\lambda(r_*)$ replaced by $\lambda(r)$.  
The magnitude of the radial flux $P^r$ is superimposed as a density plot with color-coding: negative (purple) over zero (yellow) to positive (red). 
The first quadrant of each plot in Figure \ref{Figvectorplots} corresponds to the range ($\theta \in [0 ,\pi/2]$) of Figures \ref{figscalingNHEK}-\ref{figSENHEKlogII}. When the solution is extended by symmetry to the full angular range $\theta  \in [0 , 2 \pi]$, the net energy outflux obtained by integrating the Poynting flux over the full range vanishes by energy conservation. 
To determine a non-trivial power output, we therefore consider the near horizon black hole regions embedded into an externally imposed split-monopolar magnetic field sourced by toroidal currents in a razor-thin disk at the equator \cite{Blandford:1977ds}. This magnetic field is given by the solution in the northern hemisphere which is mirrored into the southern hemisphere. All the solutions we found for near-horizon black holes threaded by a split monopole magnetic field  have net positive energy outflux. 
The AdS- and SE-NHEK solution bring in energy at the equator and emit energy near the poles. The NHEK solution has net positive energy outflux near the equator. 
Figure \ref{FigKerrvectorplots} shows the same information as Figure \ref{Figvectorplots}, but in Kerr coordinates.

\FloatBarrier



\section{Discussion and outlook}
\label{sectiondiscussion} 

We have presented symmetric FFE solutions in two different near horizon geometries of AdS-Kerr. They have the property that an observer at rest in the near horizon metric measures an energy influx, while a rotating observer -- corresponding to an observer at rest in the Kerr region -- measures an energy outflux. This behavior is consistent with the simple toy model set-up of a rotating conducting cylinder in Figure \ref{figtoymodel}, which shows that the angular flux contribution can be large enough to change the sign of net energy flux.   
The solutions have non-negative $F_\mn F^\mn \equiv 2(B^2 - E^2)$, and therefore correspond to magnetically dominated plasmas. This is required to avoid superluminal motion of the plasma. 
That said, both solutions still display an unphysical feature. The AdS-NHEK solution has an infinite energy outflux at the boundary of NHEK, similar to the NHEK scaling solution of \cite{Zhang:2014pla}.  
The SE-NHEK solution has diverging field strength near the north pole. However, as the north pole is effectively removed from the SE-NHEK geometry, this drawback might actually be irrelevant. 

We have shown the log ansatz can be used to produce an FFE solution with finite energy outflux. It would be interesting to extend this to other near-horizon geometries, most notably the original NHEK metric, which is most relevant for astrophysics. Perhaps another set of physical boundary conditions than the ones used in this paper can be used. 

The energy-extracting solutions presented in this paper require the presence of plasma, as in the usual BZ process. It could be investigated whether solutions describing a plasma-less version of the BZ process, as obtained in \cite{Jacobson:2017xam} for BTZ, also appear in the context of AdS-Kerr. 

The use of AdS geometries in this paper suggests the question whether an AdS/CFT interpretation of our FFE solutions exists, in the spirit of \cite{Jacobson:2017xam,Hawking:1998kw}. Since the AdS-Kerr solution is only known in our case in the near-horizon region, the only AdS factor that is available to us for such an AdS/CFT interpretation is the warped (at constant $\theta$) AdS$_3$ factor that makes up the 4D (AdS-)NHEK geometry. This is the AdS space that gives rise to the conjectured Kerr/CFT correspondence of \cite{Guica:2008mu}. However, an immediate repetition of the strategy in \cite{Jacobson:2017xam} for the calculation of a dual CFT conductivity would require the presence of an EM field strength component $F_{t\phi}$, which is missing by construction in the toroidally invariant solutions we presented. We leave these open questions for future work.  

\section*{Acknowledgements} 

We want to thank Elizabeth Tolman for early collaboration on this project. Part of this work was conducted at the `First USU Strings and Black Holes Workshop', which was supported by the Department of Physics and the DGCAMP group at Utah State University. The work of MJR is partially supported through the NSF grant PHY-1707571, SEV-2016-0597 and PGC2018-095976-B-C21 from MCIU/AEI/FEDER, UE. The work of N.C. is supported by the Israel Science Foundation under grant 447/17. The research of H.V. is supported by NSF grant PHY-1620059.

\appendix


\section{Kerr black holes and their near horizons} 
\label{appKerr} 

We summarize in this appendix how to take the near horizon limits of various Kerr black holes. More details can be found e.g.~in the review \cite{Compere:2012jk}. 

\subsection*{Near horizon Extreme Kerr black hole: NHEK}

In Boyer-Lindquist coordinates, the Kerr solution, parametrized by the angular momentum per mass $a$ and mass $M$, is 
\begin{eqnarray}\label{eq:Kerrmetric} 
	ds^2=-\left(1-\frac{2M{\hat{r}}}{\Sigma}\right) d{\hat{t}}^2-\frac{4Ma{\hat{r}}\sin^2{\hat{\theta}}}{\Sigma}d{\hat{t}}\,d{\hat{\phi}}+\frac{\Sigma}{\Delta}d{\hat{r}}^2+\Sigma\, d{\theta}^2+\frac{A\sin^2{\hat{\theta}}}{\Sigma} d{\hat{\phi}}^2\, 
\end{eqnarray}
where $\Sigma={\hat{r}}^2+a^2\cos^2{\hat{\theta}}$, $\Delta={\hat{r}}^2-2M{\hat{r}}+a^2$ and $A=({\hat{r}}^2+a^2)^2-\Delta\, a^2 \sin^2{\hat{\theta}}$. 
The determinant of the metric is $\sqrt{-g}=\Sigma \sin\hat{\theta}$. 
The roots of $\Delta=0$ are the black hole inner and outer event horizons $r_{\pm} = M\pm\sqrt{M^2-a^2}$. 
The Kerr black hole is stationary, axisymmetric and asymptotically flat. It has an angular velocity $\Omega \equiv -g_{\hat t \hat \phi}/g_{\hat \phi \hat \phi} = 2 M a \hat r/A$ that vanishes asymptotically. The angular velocity at the outer horizon is $\Omega_H \equiv \Omega(r_+) = a/(r_+^2 + a^2)$. 

Extremal Kerr ($a=M$) has coinciding horizons $r_+=r_-=M$ and horizon angular velocity 
\ali{
	\Omega_H^{ext} = \frac{1}{2M} \,.  \label{OmegaHextNHEK}
}
The so-called Near Horizon of Extreme Kerr (NHEK) geometry can be thought of as a zoom in on the near horizon region  $r = r_+$ of the black hole.  
Following \cite{Bardeen:1999px}, the NHEK metric \eqref{ds2NHEK} can be obtained from the  extremal Kerr black hole geometry (\ref{eq:Kerrmetric}) in Boyer-Lindquist coordinates $(\hat{t},\hat{r},\hat{\theta},\hat{\phi})$ by defining new coordinates 
\ali{
	\hat t \ra \frac{t}{\zeta} , \qquad \hat \phi \ra \phi + \frac{t}{2M\zeta}, \qquad \hat r \ra 2M^2 \zeta r + M , \qquad \hat \theta \ra \theta  \label{KerrtoNHEK}
}
and taking the limit $\zeta \rightarrow 0$. 
The location of the horizon
is at $r=0$ in these coordinates. From the relation $\phi = \hat \phi - \frac{1}{2M} \hat t$ between the new angle $\phi$ and the Kerr angle $\hat \phi$, it is clear that the NHEK geometry co-rotates with extremal Kerr at the angular velocity of the horizon \eqref{OmegaHextNHEK}. 
The scale parameter $\zeta$ can be rescaled without changing the resulting near horizon metric. This points to an extra scaling symmetry of the NHEK geometry under $r \ra \lambda r $ and $t \ra t/\lambda$. It is discussed in more detail in section \ref{sectionNHEK}. 


\subsection*{Near horizon extreme AdS-Kerr black hole: AdS-NHEK} \label{appAdSNHEK}

In this section and the next we discuss two types of near horizon geometries that follow from two types of extremal limits of the AdS-Kerr black hole (in Boyer-Lindquist coordinates) 
\begin{eqnarray}\label{KerrAds}
ds^2=-\frac{\Delta_a}{\Sigma_a}\left(d\hat{t}-\frac{a\sin^2\hat{\theta}}{\Xi}\, d\hat{\phi} \right)^2+\frac{\Sigma_a}{\Delta_a}\,d\hat{r}^2+\frac{\Sigma_a}{\Delta_{\theta}} \,d\hat{\theta}^2+\frac{\Delta_{\theta} \sin^2\hat{\theta}}{\Sigma_a}\left(a\,d\hat{t}-\frac{\hat{r}^2+a^2}{\Xi} \,d\hat{\phi} \right)^2
\end{eqnarray}
where
\begin{eqnarray}
\Delta_a=(\hat{r}^2+a^2)(1+\hat{r}^2/l^2)-2m \hat{r}\,\qquad\Xi=1-a^2/l^2\,,\\
\Delta_{\theta}=1-a^2 \cos^2\hat{\theta}/l^2\,\qquad \Sigma_a=\hat{r}^2+a^2\cos^2\hat{\theta} \, . 
\end{eqnarray}
In one case we take the traditional extreme limit where the inner and outer horizons degenerate to a single horizon at $\hat r=r_+=r_-$. In the other case we will take the so called super-entropic limit $a \rightarrow l$ while also fixing the mass $m=8\, l/3\sqrt{3}$ for the black hole to be extreme. More details on these two different prescriptions are described next. 

The outer and inner horizon of AdS-Kerr are defined as the largest and smallest root of $\Delta_a$ (which allows to eliminate $m$ for $r_\pm$).    
As it was shown in  \cite{Hartman:2008pb,Compere:2012jk}, to find the near horizon geometry of the extreme AdS-Kerr black hole (\ref{NHE}) one has to first find the extreme limit of $(\ref{KerrAds})$ where the inner and outer horizons degenerate  $r=r_+=r_-$, then introduce new coordinates
\ali{
	\hat t \ra t \frac{r_0}{\epsilon},  \qquad \hat \phi \ra \phi + \Omega_H^{ext} \frac{t \, r_0}{\epsilon}, \qquad \hat r \ra r_+ + \epsilon \, r_0 \, r ,  \qquad \hat \theta \ra \theta ,  \label{AdSNHEKlimit}
}
with $r_0$ and $\Omega_H^{ext}$ defined below, and finally take the limit $\epsilon \rightarrow 0$. 

The extremality condition $a(r_+)$ is given in $(\ref{eq:horizon})$ and $r_0$ is defined as 
\begin{eqnarray}
r_0^2=\frac{(r_+^2 +a^2)}{\Delta_0}\, , \qquad \Delta_0 = 1 + \frac{a^2}{l^2} + 6 \frac{r_+^2}{l^2} \, .  \label{defr0}
\end{eqnarray}
The parameter $\Omega_H^{ext}$ is given by the AdS-Kerr angular velocity of the (outer) horizon 
\begin{eqnarray}
\qquad \Omega^{ext}_H = \frac{\Xi\, a}{r_+^2 +a^2}\, ,   \label{OmegaHextAdS}
\end{eqnarray}
where the superscript refers to the angular velocity being evaluated at extremality $(\ref{eq:horizon})$. 

\subsection*{Near horizon of super-entropic AdS-Kerr black hole: SE-NHEK}
\label{subsec:SENEHK}

A different extremal limit for the AdS-Kerr black hole has been  studied in the super-entropic limit $a \rightarrow l$ in  \cite{Hennigar:2015cja}. Here we show how to derive the near horizon geometry of these  super-entropic extremal black holes, as first discussed in  \cite{Sinamuli:2015drn}.  \\  

Following \cite{Hennigar:2015cja} we first take the limit $a\rightarrow l$, but at the same time to avoid a singularity in the metric rescale the coordinate $\hat{\phi}=\hat{\psi} \,\Xi$ and identify $\hat \psi$ with period $2\pi/\Xi$  to prevent conical singularities. The black hole (\ref{KerrAds}) in this limit is super-entropic (see \cite{Cvetic:2010jb} for details) and becomes 
\begin{eqnarray}\label{eq:SEKerr}
ds^2=-\frac{\Delta_l}{\Sigma_l}\left(d\hat{t}-{l\sin^2\hat{\theta}}\, d\hat{\psi} \right)^2+\frac{\Sigma_l}{\Delta_l}\,d\hat{r}^2+\frac{\Sigma_l}{\sin^2\hat{\theta}} \,d\hat{\theta}^2+\frac{ \sin^4\hat{\theta}}{\Sigma_l}\left(l\,d\hat{t}-(\hat{r}^2+l^2) \,d\hat{\psi} \right)^2
\end{eqnarray}
where
\begin{eqnarray}
\Delta_l=(\hat{r}^2+l^2)(1+\hat{r}^2/l^2)-2m \hat{r}\,\qquad\Sigma_l=\hat{r}^2+l^2\cos^2\hat{\theta} \, . 
\end{eqnarray}
The coordinate $\hat \psi$ is non-compact and hence we choose to compactify it by $\hat{\psi}\sim\hat{\psi}+2\pi$.
The black hole event horizon topology is that of a sphere with two punctures (at the poles) and for horizons to exist there is a minimal extremal value of the mass $m\equiv m_l=8\,l/(3\sqrt{3})$. Note that now there are only three roots of $\Delta_l |_{m=m_l} =0$ located at $r=\{l/\sqrt{3}, (-1\pm  2 \sqrt{2} i) \,l/\sqrt{3}\}$. The event horizon is therefore located at 
\ali{ 
	r_+ \equiv r_l=\frac{l}{\sqrt{3}} . \label{rpSE}
	}  

Now we derive the near horizon region $r=r_l$ of these extreme super-entropic black holes, dubbed here SE-NHEK metric  \eqref{NHE} with functions $(\ref{eq:SEdefd})$. It can be obtained from the extremal $m=m_l$ super-entropic AdS-Kerr black hole geometry (\ref{eq:SEKerr}) by defining new coordinates in the same way as in \eqref{AdSNHEKlimit} but starting from the super-entropic black hole \eqref{eq:SEKerr} with angle $\hat \psi$, namely 
\ali{
	\hat t \ra t \frac{r_0}{\epsilon},  \qquad \hat \psi \ra \phi + \Omega_H^{ext} \frac{t \, r_0}{\epsilon}, \qquad \hat r \ra r_+ + \epsilon \, r_0 \, r ,  \qquad \hat \theta \ra \theta  ,  \label{SENHEKlimit}
}
and taking the limit $\epsilon \rightarrow0$. 
Note that in this super-entropic extremal limit $r_0$ defined in \eqref{defr0} equals $r_+$, i.e. 
\ali{
	r_0 
	= \frac{l}{\sqrt{3}} \, ,  \label{r0SE}
}
and $\Omega_H^{ext}$ is given by the horizon angular velocity $\Omega_H \equiv -g_{\hat t \hat \psi}/g_{\hat \psi \hat \psi} (r_+)$ of the  geometry \eqref{eq:SEKerr}, evaluated at the extremal value of $r_+$ in \eqref{rpSE}, 
\ali{
	\Omega_H^{ext} = \frac{3}{4 l}\,. \label{OmegaHSE}
}
With these values filled in, the SE-NHEK limit \eqref{SENHEKlimit} can be written out as 
\begin{eqnarray}
\hat{t}\rightarrow r_l\,t/\epsilon\,, \qquad\hat{\psi}\rightarrow \phi+(\sqrt{3}/4)\,t/\epsilon \, , \qquad\hat{r}\rightarrow r_l\,(1+\epsilon\, r)   \,,\qquad\hat{\theta}\rightarrow \theta\,.
\end{eqnarray}

It was shown in \cite{Sinamuli:2015drn} that the super-entropic limit and the near-horizon limit of AdS-Kerr commute, i.e.~the SE-NHEK metric can alternatively be obtained from a super-entropic limit of AdS-NHEK.


\subsection*{Near horizon of extremal BTZ: NHEBTZ} \label{appBTZ}

The BTZ black hole background metric \cite{Banados:1992wn}, which is a three-dimensional solution of Einstein's equations with a negative cosmological constant, is   
\ali{
	ds^2 &= -\alpha(\hat r)^2 d\hat t^2 + \frac{d\hat r^2}{\alpha(\hat r)^2} + \hat r^2 ( d\hat \phi - \Omega(\hat r) d\hat t)^2   \label{ds2BTZ}
}
with 
\ali{
	\alpha(\hat r) = \frac{(\hat r^2-r_+^2)(\hat r^2-r_-^2)}{\hat r^2 l^2}, \qquad \Omega(\hat r) = \frac{r_- r_+}{\hat r^2 l} .   \label{alphaOmegaBTZ}
}
It has horizons $r_\pm$ at the locations where $\alpha(\hat r)$ vanishes, and angular velocity $\Omega(\hat r)$. 
In the extremal limit, the inner and outer horizon coincide $r_+=r_-$ and the horizon angular velocity becomes 
\ali{
	\Omega_H^{ext} = \frac{1}{l} \, .  \label{OmegaHextBTZ}
} 
The near-horizon extreme BTZ geometry or NHEBTZ geometry \eqref{ds2nhBTZ} 
can be obtained from the coordinate transformation (see e.g.~\cite{deBoer:2010ac}) 
\ali{
	\hat  t = \frac{t}{\epsilon},  \qquad \hat \phi = \phi + \frac{t}{\epsilon l} , \qquad \hat r^2 = r_+^2 + \epsilon r  \label{nhBTZcoord}
} 
in the limit $\epsilon \ra 0$.

\section{Stream equations in near-horizon regions} \label{appEOM} 

For the NHEK metric (\ref{NHE}) with \eqref{ds2NHEK}, the action in \eqref{S2P} becomes 
\ali{
	\begin{split} 
		\mathcal{S}[A_\phi] &= \int dr d\theta \left[ \frac{d}{2} \left(r^2 (\p_r A_\phi)^2 + (\p_\theta A_\phi)^2 \right) + \frac{1}{2 r^2 \Lambda} I(A_\phi)^2 \right] \\
		& \qquad \qquad d \equiv -\frac{\mathcal C}{r} = \Lambda (\frac{\omega_F}{r}+1)^2 -\frac{1}{\Lambda} 
	\end{split}   \label{S2PNHEK}
}
or alternatively, using \eqref{dAtofdAphi}, 
\ali{
	\begin{split} 
		\mathcal{S}[A_t] &= \int \frac{dr d\theta}{r^2} \left[  \frac{D}{2} \left(r^2 (\p_r A_t)^2 + (\p_\theta A_t)^2 \right) + \frac{1}{2 \Lambda} I(A_t)^2 \right] \\
		& \qquad \qquad D \equiv -\frac{r}{\omega_F^2} \mathcal C = \Lambda \left(1 + \frac{r}{\omega_F}\right)^2 -\frac{1}{\Lambda} \frac{r^2}{\omega_F^2} . 
	\end{split}  
}
From these actions we obtain respectively the NHEK EOM for $A_\phi$, 
\ali{
	\p_\theta ( d \, \p_\theta A_\phi) + \p_r ( d \, r^2 \p_r A_\phi) - \frac{1}{2}  \frac{\delta d}{\delta A_\phi} \left( r^2 (\p_r A_\phi)^2 + (\p_\theta A_\phi)^2 \right) -  \frac{1}{r^2 \Lambda} I \frac{\delta I}{\delta A_\phi} = 0, \label{AphiEOMNHEK}
}
and the one for $A_t$, 
\ali{
	\p_\theta ( \frac{D}{r^2} \p_\theta A_t) + \p_r ( D \p_r A_t) - \frac{1}{2} \frac{\delta D}{\delta A_t}  \left( (\p_r A_t)^2 + \frac{1}{r^2}(\p_\theta A_t)^2 \right) -  \frac{1}{r^2 \Lambda} I \frac{\delta I}{\delta A_t} = 0. \label{AtEOMNHEK}
} 

For an AdS-NHEK metric \eqref{NHE}, with either (\ref{eq:AdsDefs}) or (\ref{eq:SEdefd}), the action \eqref{S2P} is  
\ali{
	\begin{split} 
		\mathcal{S}[A_\phi] &= \int dr d\theta \left[ \frac{d_\alpha}{2} \left(r^2 (\p_r A_\phi)^2 + \frac{1}{\alpha^2} (\p_\theta A_\phi)^2 \right) +  \frac{\alpha}{2 r^2 \gamma} I(A_\phi)^2 \right]    \\
		&\qquad \qquad  d_\alpha \equiv -\frac{\alpha}{r}\mathcal C = \alpha \gamma (\frac{\omega_F}{r}+k)^2 -\frac{\alpha}{\gamma} 
	\end{split} \label{eq157}
}
or
\ali{
	\begin{split} 
		\mathcal{S}[A_t] &= \int \frac{dr d\theta}{r^2} \left[  \frac{D_\alpha}{2} \left(r^2 (\p_r A_t)^2 + \frac{1}{\alpha^2}(\p_\theta A_t)^2 \right) +  \frac{\alpha}{2 \gamma} I(A_t)^2 \right] \\
		& \qquad \qquad D_\alpha \equiv -\frac{\alpha r}{\omega_F^2} \mathcal C = \gamma \alpha \left(1 + k \frac{r}{\omega_F}\right)^2 -\frac{\alpha}{\gamma} \frac{r^2}{\omega_F^2}  \, . 
	\end{split} 
} 
Variation of the action gives the AdS-NHEK EOM for $A_\phi$, 
\ali{
	\p_\theta ( d_\alpha \, \frac{1}{\alpha^2} \p_\theta A_\phi) + \p_r ( d_\alpha \, r^2 \p_r A_\phi) - \frac{1}{2}  \frac{\delta d_\alpha}{\delta A_\phi} \left( r^2 (\p_r A_\phi)^2 + \frac{1}{\alpha^2}(\p_\theta A_\phi)^2 \right) - \frac{ \alpha}{r^2 \gamma} I \frac{\delta I}{\delta A_\phi} = 0,   \label{AphiEOMAdS}
}
and for $A_t$, 
\ali{
	\p_\theta ( \frac{D_\alpha}{r^2 \alpha^2} \p_\theta A_t) + \p_r ( D_\alpha \p_r A_t) - \frac{1}{2} \frac{\delta D_\alpha}{\delta A_t}  \left( (\p_r A_t)^2 + \frac{1}{r^2 \alpha^2}(\p_\theta A_t)^2 \right) - \frac{\alpha}{r^2 \gamma} I \frac{\delta I}{\delta A_t} = 0. \label{AtEOMAdS}  
}

\section{EM fields in different frames} \label{appEM} 

We review the identification of EM fields in the EM tensor that allow the standard form of Maxwell's equations to be obtained from the covariant expressions in terms of the field strength. Different frames are considered. In particular the relation between electromagnetic fields in rotating frames in \eqref{Erot} is useful for the description of the toy model in the main text.  

\paragraph{Carthesian} In Carthesian coordinates in mostly plus  convention, the standard definition 
\ali{
	F_\mn = \left(\begin{array}{cccc} 0 & -E_x & -E_y & -E_z \\ E_x & 0 & B_z & -B_y \\ E_y & -B_z & 0 & B_x \\ E_z & B_y & -B_x & 0  
	\end{array} \right)  \label{Fcarthesian}
}
(or $E_i = F_{i0}$ and $B_i = \frac{1}{2} \epsilon_{0ijk} F^{jk}$, with $\epsilon_{\mn \alpha\beta}$ the covariant permutation tensor of the spacetime) 
is easily shown to provide as equivalent formulations of Maxwell's equations 
\ali{
	\p_\nu F^\mn = j^\mu \qquad \qquad \left\{ \begin{array}{ll}  \nabla \cdot E = j^0 \\  \nabla \times  B = \frac{\p  E}{\p t} +  j
	\end{array} \right. \label{Maxwell}
}
and 
\ali{
	\p_\nu \tilde F^\mn = 0 \qquad \qquad \left\{ \begin{array}{ll}  \nabla \cdot  B = 0 \\  \nabla \times E = -\frac{\p B}{\p t} 
	\end{array} \right. , \label{Maxwelltilde}
}	
with 
\ali{ 
	\tilde F^\mn = \frac{1}{2} \epsilon^{\mn  \alpha \beta} F_{\alpha \beta}.   
} 

The electromagnetic stress energy tensor is defined as 
\ali{
	T_\mn = F_{\mu \alpha} F_\nu^{\phantom{\nu}\alpha} - \frac{1}{4} F_{\alpha\beta} F^{\alpha\beta} g_\mn 
}
and related to the Poynting vector $P^i = (E \times B)^i = \epsilon_0^{\phantom{0}ijk} E_j B_k$ as  
\ali{
	T^i_{\phantom{ii}0} = - P^i.  
	\label{TmnofPoynting}
}

\paragraph{Cylindrical} In cylindrical coordinates, we use 
\ali{
	F_\mn &= \left(\begin{array}{cccc} 0 & -E_r & -E_\theta & -E_z \\ E_r & 0 & B_z r & -B_\theta/r \\ E_\theta & -B_z r & 0 & B_r r \\ E_z & B_\theta/r & -B_r r & 0  
	\end{array} \right)	\label{Fcylinder}
}
to obtain the equivalent formulations 
\ali{
	\nabla_\nu F^\mn = j^\mu \qquad \qquad \left\{ \begin{array}{ll} \nabla \cdot E = j^0 \\ \nabla \times B = \frac{\p  E}{\p t} 
		+  j
	\end{array} \right. \label{Maxwellcyl}
}
and 
\ali{
	\nabla_\nu \tilde F^\mn = 0 \qquad \qquad \left\{ \begin{array}{ll}  \nabla \cdot B = 0 \\ \nabla \times E = - \frac{\p  B}{\p t} 
	\end{array} \right. . \label{Maxwelltildecyl}
}
Here, $\nabla_\mu$ is the covariant derivative in the metric 
\ali{
	ds^2 = -dt^2 + dr^2 + r^2 d\theta^2 + dz^2 .  \label{ds2cyl}
}
The field strength \eqref{Fcylinder} can be obtained from the Carthesian expression \eqref{Fcarthesian} by the transformation $F \ra \Lambda^T F \Lambda$ under the transformation from Carthesian coordinates $x$ to cylindrical coordinates $x'$ with transformation matrix $\Lambda = \p x/\p x'$. The components of the electromagnetic fields are given by $E_i = F_{i0}$ and $B_i = \frac{1}{2} \epsilon_{0ijk} F^{jk}$. In the alternative choice of basis $\vec e_\theta \ra \sqrt{g_{\theta \theta}} \vec e_\theta$ (or $E_\theta = E_\theta^{cyl}/r$ and $B_\theta = B_\theta^{cyl}/r$), the form of Maxwell's equations in curvilinear, cylindrical coordinates is recovered. 

The same relation \eqref{TmnofPoynting} holds for the Poynting flux defined as $P^i = (E \times B)^i = \epsilon_0^{\phantom{0}ijk} E_j B_k$,  
\ali{
	P = \left( \frac{B_z E_\theta-B_\theta E_z}{r},  \frac{B_r E_z-B_z E_r}{r} ,  \frac{B_\theta E_r - B_r E_\theta}{r} \right).   \label{Plabcyl}
}

\paragraph{Rotating frame} We can go from cylindrical coordinates $x^\mu = (t,r,\theta,z) = (t',r',\theta'+\omega t',z')$ to rotating cylindrical coordinates $x^{\mu'} = (t', r', \theta', z')=(t,r,\theta - \omega t,z)$, with metric $ds^2 = -dt^2 + dr^2 + r^2 d\theta^2 + dz^2$ transformed into the metric $ds'^2 = -dt'^2 + dr'^2 + r'^2 (d\theta' + \omega dt')^2 + dz'^2$ such that the primed frame rotates with angular velocity $\omega$ counterclockwise about the $z$ axis.   
In the rotating frame in cylindrical coordinates $(t',r',\theta',z')$, 
the transformed field strength is then given by\footnote{
	One goes between contravariant and covariant EM fields by $E^\theta = E_\theta/r^2$ and $B^\theta = B_\theta/r^2$. Also, as remarked in the previous section on the cylindrical frame, we can use $E^\theta \ra E^\theta/r$, $B^\theta \ra B^\theta/r$ to recover the curvilinear notation.   
}  
\ali{
	F^{\mu'\nu'}_{rot} &= \left(\begin{array}{cccc} 0 & E_r &  E_\theta/r^2 &  E_z \\  -E_r & 0 & B_z^{rot}/r & -B_\theta^{rot}/r \\  -E_\theta/r^2 & -B_z^{rot}/r & 0 & B_r^{rot}/r \\ -E_z & B_\theta^{rot}/r & -B_r^{rot}/r & 0  
	\end{array} \right) 
	\label{rotatingFmn}
}
and 
\ali{
	F_{\mu'\nu'}^{rot} &= \left(\begin{array}{cccc} 0 & -E_r^{rot} & -E_\theta^{rot} & -E_z^{rot} \\ E_r^{rot} & 0 & r B_z & -B_\theta/r \\ E_\theta^{rot} & -r B_z & 0 & r B_r \\ E_z^{rot} & B_\theta/r & -r B_r & 0  \end{array} \right)	
}
with the rotating EM fields as a function of the lab EM fields given by 
\ali{
	E_\theta^{rot} = E_\theta, \quad  E_r^{rot} = E_r + \omega r B_z, \quad E_z^{rot} = E_z - \omega r B_r   \label{Erot}\\ 
	B_\theta^{rot} = B_\theta, \quad B_r^{rot} = B_r - \omega r E_z, \quad B_z^{rot} = B_z + \omega r E_r \label{Brot}
}
or $E_{rot} = E + (\omega \vec e_z \times \vec r) \times B$ and $B_{rot} = B - (\omega \vec e_z  \times \vec r) \times E$. 
The rotating fields are obtained from the field strength through 
\ali{
	E_{i'}^{rot} = F_{i'0'} , \qquad  B_{i'}^{rot} = \frac{1}{2} \epsilon_{0'i'j'k'} F^{j'k'} . 	
} 

The field strength in \eqref{rotatingFmn} gives the Maxwell equations 
\ali{
	\nabla_{\nu'} F^{\mu'\nu'}_{rot} = j^{\mu'}_{rot} \qquad \qquad \left\{ \begin{array}{ll} \nabla \cdot E = j^{0'}_{rot} \\ \nabla \times B_{rot} = \frac{\p E}{\p t} + j_{rot}
	\end{array} \right.   \label{rotatingmaxwell}
}
and 
\ali{
	\nabla_{\nu'} \tilde F^{\mu'\nu'}_{rot} = 0 \qquad \qquad \left\{ \begin{array}{ll} \nabla \cdot B = 0 \\ \nabla \times E_{rot} = -\frac{\p B}{\p t} 
	\end{array} \right. \, ,   \label{arotatingmaxwelleq}
}	
with 
\ali{
	j^{\mu'}_{rot} = \left( \rho, j^r, j^\theta - \rho \, \omega, j^z \right)   
}
and $\nabla_{\mu'}$ the covariant derivative in the metric 
\ali{ 
	ds'^2 = -dt'^2 + dr'^2 + r'^2 (d\theta' + \omega dt')^2 + dz'^2. \label{rotcylmetric}
} 
The rotating frame Poynting flux 
\ali{
	P_{rot} &= \left( \frac{B_z^{rot} E_\theta^{rot}-B_\theta^{rot} E_z^{rot}}{r'},  \frac{B_r^{rot} E_z^{rot}-B_z^{rot} E_r^{rot}}{r'} ,  \frac{B_\theta^{rot} E_r^{rot} - B_r^{rot} E_\theta^{rot}}{r'} \right)   \label{Prot}
}
is determined by the Maxwell stress tensor through 
\ali{
	T^{i'}_{\phantom{ii}0'} &= - P^{i'}_{rot} \,\, .  \label{ProtfromT}
} 

Further references on EM fields in rotating frames are \cite{Arendt:1998vq} and \cite{ThorneMacDonald}.  


\section{General curved space EM field definitions} \label{appCurvedEM}

For $u$ a future-pointing unit time-like vector field ($u^\mu u_\mu = -1$) in a 4-dimensional curved spacetime $g_\mn$, 
the associated observer with four-velocity $u$ measures EM fields (see e.g. \cite{Felice:2010cra}) 
\ali{
	E_\mu &= F_\mn u^\nu   \label{Ecurvedspacedef} \\
	B_\mu &= \frac{1}{2} \epsilon_{\sigma \mu \alpha \beta} F^{\alpha \beta} u^\sigma  \label{Bcurvedspacedef}
}
and Poynting vector 
\ali{
	P^\mu = \epsilon^{\sigma \mu \nu \rho} u_\sigma E_\nu B_\rho.  \label{Pcurvedspacedef}  
}
The Poynting vector is defined as  
\ali{
	P^\mu = -\gamma^\mu_{\phantom{\mu}\alpha}T^{\alpha \beta} u_\beta =  -T^\mu_{\phantom{\alpha}\rho} u^\rho - u^\mu u_\beta T^\beta_{\phantom{\beta}\rho} u^\rho   \label{Pofudef}
}
in terms of the induced metric $\gamma_{\alpha\beta}$ on the hypersurface orthogonal to $u$, 
\ali{
	\gamma^{\alpha\beta} = g^{\alpha\beta} + u^\alpha u^\beta \,\, . 
}

\paragraph{Rotating cylinder }  
We now apply these definitions to the rotating cylinder spacetime \eqref{rotcylmetric}. For an observer $u^{\mu'} = (1,0,0,0)$ at rest in the rotating frame, \eqref{Ecurvedspacedef} and \eqref{Bcurvedspacedef} extract the fields $E = (E_r^{rot},E_\theta^{rot},E_z^{rot})$ and $B = (B_r^{rot},B_\theta^{rot},B_z^{rot})$, and \eqref{Pofudef} gives \eqref{ProtfromT} for the Poynting flux \eqref{Prot}. (Notice however that \eqref{Pcurvedspacedef} does not reproduce \eqref{Prot}, because the observer's four-velocity is not normalized to unity.)

On the other hand, the unit normal to a constant time surface is given by $u_{\mu'} = (-1,0,0,0)$. The corresponding observer 
\ali{
	u^{\mu'} = (1,0,-\omega,0)  \label{eqB6}
}
is a zero angular momentum observer or `ZAMO', as it satisfies $u^{\mu'} \eta_{\mu'} = 0$ with 
$\eta$ the angular Killing vector of the spacetime. 
The definitions \eqref{Ecurvedspacedef}-\eqref{Pcurvedspacedef} then extract the lab EM fields $E = (E_r,E_\theta,E_z)$ and $B = (B_r,B_\theta,B_z)$, as well as the the lab Poynting vector \eqref{Plabcyl}.  
Indeed, the ZAMO rotates in the rotating frame and corresponds to an observer at rest in the lab frame \eqref{ds2cyl}, 
\ali{
	u^\mu = (1,0,0,0) \label{eqB7}
} 
via	$u^{\mu'} = \frac{\p x^{\mu'}}{\p x^\mu} u^\mu$. \\


\bibliographystyle{JHEP}
\bibliography{referencesFFE}

\end{document}